\documentclass[11pt,a4paper]{article}
\usepackage[left=23mm,right=23mm,top=25mm,bottom=25mm,headheight=14pt,headsep=8mm]{geometry}
\usepackage{iftex}
\ifPDFTeX
  \usepackage[T1]{fontenc}
  \usepackage[utf8]{inputenc}
  \usepackage{lmodern}
\else
  \usepackage{fontspec}
\fi
\usepackage{amsmath,amssymb,amsthm,mathtools}
\usepackage{booktabs,tabularx,array,graphicx,xcolor}
\usepackage{enumitem,fancyhdr,setspace,caption,float}
\usepackage[protrusion=true,expansion=false]{microtype}
\usepackage{titlesec,placeins}
\usepackage[hidelinks,unicode]{hyperref}
\hypersetup{pdftitle={Interference-Free Capacity of a Binary Interference Channel with Causal Observation},pdfauthor={Xianwei Meng},pdfsubject={R4 single-file source release, 2026-09-18}}
\setlist{nosep,leftmargin=2em}

\titleformat{\section}{\normalfont\large\bfseries}{\thesection.}{0.65em}{}
\titleformat{\subsection}{\normalfont\normalsize\bfseries}{\thesubsection.}{0.65em}{}
\titlespacing*{\section}{0pt}{2.1ex plus .5ex minus .2ex}{1.0ex plus .2ex}
\titlespacing*{\subsection}{0pt}{1.8ex plus .4ex minus .2ex}{.7ex plus .2ex}
\newcommand{\E}{\mathbb E}
\newcommand{\ind}{\mathbf 1}
\newcommand{\rank}{\operatorname{rank}}
\begin{document}
\thispagestyle{plain}
\begin{center}
{\fontsize{18}{22}\selectfont\bfseries
Interference-Free Capacity of a Binary\\[2pt]
Interference Channel with Causal Observation\par}
\vspace{10pt}
{\normalsize Xianwei Meng\textsuperscript{*}\par}
\vspace{3pt}
{\small Hefei University of Technology\par}
\vspace{3pt}
{\small \textsuperscript{*}Corresponding author: \href{mailto:mxianwei@hfut.edu.cn}{mxianwei@hfut.edu.cn}\par}
\end{center}
\vspace{2pt}
\begin{abstract}
\noindent
We determine the capacity region of a two-user binary interference channel with receiver-controlled observation and prove that causal port selection attains the interference-free limit. The physical--observation dual-axis formulation makes the observation kernel a design variable subject to explicit causal and resource constraints. For two equiprobable, initially unknown block states, matching converses and coding constructions give exact sum capacities of 1, $4/3$, and 2 bits per channel use for fixed-port, open-loop, and causal policies, respectively. These results hold for every fixed joint error threshold below $1/2$, under both average and maximal message error, with error averaged over the block state. Every receiver makes one observation per slot and switches ports at most once; the transmitters receive no feedback and the receivers do not cooperate. Causal observation thus doubles the optimal fixed-port sum capacity and exceeds the optimal open-loop sum capacity by 50\%. With independent erasures of retention probability $p$, the exact causal region is $[0,p]^2$. A finite-blocklength bound accounts jointly for state-identification pilots and coding redundancy. We also optimize observation policies for fixed codes: an exact finite-horizon solution reduces message error from 10.76\% to 7.52\%, and a second-order approximation bound controls optimization over continuous observation directions. Linear-array designs and coding experiments examine the corresponding training, reliability, and control costs. The capacity result identifies a setting in which observation design removes the entire interference penalty, with the gain established by a converse as well as an achievable construction.
\end{abstract}
\begingroup\small
\par
\endgroup

\section{Introduction}

Determining the rates at which independent users can communicate over a shared channel has been a central problem of information theory for decades. The Han--Kobayashi achievable region~\cite{HanKobayashi1981} and the one-bit approximation for the two-user Gaussian interference channel~\cite{EtkinTseWang2008} mark two stages in this development. Both address coding and decoding for a prescribed channel law. The ability to choose how the next observation is formed makes the admissible observation policies part of the capacity problem. This paper gives an exact solution for a binary interference channel whose receivers can select between two physical ports.

The physical--observation dual-axis (PODA) structure distinguishes the physical state from the conditions under which a record is formed~\cite{MengDual2026}. In communication, this places the transmitted symbols and propagation environment on the physical axis, and the selectable formation kernel on the observation axis. The distinction concerns the roles of the variables and does not require statistical independence. An observation action can depend on the receiver's past record, so a protocol specifies both a code and a causal policy. Its performance must be evaluated under their joint probability law, with the cost of acquiring the record included.

Our main result determines all three capacity regions obtained from fixed ports, predetermined port sequences, and causal port selection in the same binary port family. With two equiprobable block states and any fixed joint error threshold below $1/2$, their sum capacities are, respectively, $1$, $4/3$, and $2$ bits per channel use. The last value is the sum capacity of two interference-free binary links. Causal observation therefore recovers the entire interference-free region, doubles the optimum over fixed ports, and improves on the optimum over all open-loop policies by 50\%. The comparison is between capacity regions, with matching converse bounds and coding constructions. It holds under average and maximal message error, where the environmental state is averaged in both cases. All constructions use one observation per receiver per slot and at most one port switch, without transmitter feedback, message sharing, or receiver cooperation.

The mechanism is explicit. The two ports produce either the desired symbol or the sum modulo two of the users' symbols; their roles are determined by an unknown state that remains fixed during the block. One known pilot identifies the clean port. A causal receiver uses that information to select its subsequent observations, whereas an open-loop receiver can use it only in decoding. The open-loop converse couples correct decoding in the two states and bounds the distinguishable message pairs. A three-phase code attains that bound with one switch. The causal construction attains the output-history counting bound. Thus the exact separation measures the communication value of using acquired information to control later observations.

With independent erasures, causal observation still attains the interference-free region $[0,p]^2$, where $p$ is the retention probability. State identification uses a sublinear number of pilots, and random linear coding supplies the required redundancy. At finite blocklength, the pilot length and payload are optimized together. The existence of a clean channel section, its identification from known pilots, and the negligible asymptotic identification cost are the structural conditions behind this capacity equality. The result solves the stated receiver-controlled binary model; the capacity region of a general prescribed interference channel is a different coding problem. Even within the erasure extension considered here, the fixed-port and open-loop regions are only bounded.

Observation design also matters before taking a capacity limit. For a fixed code, the joint posterior of messages and environment, together with the remaining budget, gives a finite-horizon decision recursion. In a five-observation model, the exact optimum reduces average message error from 10.76\% for predetermined observations to 7.52\% for causal policies. For continuous directions, we bound the gap between finite-grid and continuous optima. The optimized policies also show that maximizing task mutual information and minimizing decision error need not select the same observations.

Finally, a linear array with a single sampling chain provides a continuous implementation of observation formation. We derive combining weights, robust designs, and prediction bounds under explicit training assumptions, then compare actual coded transmissions under common budgets. These results distinguish the existence of a useful observation direction from the cost of learning and tracking it. Together, the exact capacity theorem and the finite-code designs establish how control of the observation axis can remove an interference penalty in a specified channel and improve message recovery when observation resources are limited.

\section{Formation Kernels and Causal Policies}
\subsection{Dual-Axis Structure and Protocol-Induced Channels}

Let $\mathcal X$, $\mathcal S$, and $\mathcal Y$ be standard Borel spaces, and let the admissible domain $\mathcal B\subseteq\mathcal X\times\mathcal S$ be measurable. An admissible physical state $x$ and observation state $s$ form a record through the kernel
\begin{equation}
 K(\mathrm dy\mid x,s),\qquad (x,s)\in\mathcal B
 \label{eq:formation}
\end{equation}
Here $\mathcal B$ specifies admissibility across the two axes, and $Y$ is the resulting record. Fixing $s=s_0$ gives the channel section $W_{s_0}(\mathrm dy\mid x)=K(\mathrm dy\mid x,s_0)$; varying $s$ may change the response, noise, distinguishable directions, or feasible domain.

For a protocol $\mathsf p$, let $\Lambda_{\mathsf p}(\mathrm dx,\mathrm ds\mid u)$ be its realization kernel, supported on $\mathcal B$. Let $L_{\mathsf p}(\mathrm dr\mid s,y)$ be the fine-record kernel and $T_{\mathsf p}(\mathrm dz\mid r)$ the readout kernel. Then
\begin{align}
 W_{\mathsf p}^{R}(\mathrm dr\mid u)
 &=\int_{\mathcal B}\int_{\mathcal Y}
 L_{\mathsf p}(\mathrm dr\mid s,y)K(\mathrm dy\mid x,s)
 \Lambda_{\mathsf p}(\mathrm dx,\mathrm ds\mid u),\\
 W_{\mathsf p}^{Z}(\mathrm dz\mid u)
 &=\int T_{\mathsf p}(\mathrm dz\mid r)
 W_{\mathsf p}^{R}(\mathrm dr\mid u).
 \label{eq:strategy}
\end{align}
The fine record may include the visible observation state $S$, giving the chain $U\to(X,S)\to(S,Y)\to R\to Z$. Discarding $S$ from an acquired record is a readout operation; changing $S$ selects another formation kernel. Equation~\eqref{eq:strategy} describes a single stage. A causal block channel is obtained by composing the successive conditional kernels.

Admissibility of $\Lambda_{\mathsf p}$ is determined by each terminal's information, control permissions, and causal order. Maps on finite alphabets are automatically measurable. The continuous array models use explicit matrix operations and a fixed rule for breaking ties. Standard Borel spaces provide a common formulation; the optimality and capacity results below are established for their respective specified models.

\subsection{Conditional Mutual Information and Observation Actions}

When the relevant mutual informations are finite, the chain rule gives
\begin{align}
 I(X,S;Z)&=I(X;Z)+I(S;Z\mid X)\\
         &=I(S;Z)+I(X;Z\mid S).
 \label{eq:two-axis-chain}
\end{align}
These are two expansions of the same joint information. Each conditional term depends on the other axis; comparisons and optimizations must retain the probability law and resource constraints induced by a common policy.

The receivers considered here have no additional messages to transmit. Write their local histories as $\mathcal H_{i,t-1}=(S_i^{t-1},Y_i^{t-1})$ and their local random seeds as $V_i$. The observation actions satisfy
\begin{equation}
 S_{i,t}=\pi_{i,t}(V_i,\mathcal H_{i,t-1}),\qquad
 I(M_1,M_2;S_{i,t}\mid V_i,\mathcal H_{i,t-1})=0.
 \label{eq:local-policy}
\end{equation}
Any dependence between the action and the messages arises through the available history. Conditional on that history, the action selects the law of the next observation. Its communication value is measured by the resulting ability to distinguish messages.

For an acquired fine record $R$ and a parameter-independent readout $R\to Z$,
\begin{equation}
 I(M;R)=I(M;Z)+I(M;R\mid Z).
 \label{eq:record-loss}
\end{equation}
A receiver with the finer record can apply the original readout kernel and decoder, reproducing the same error probability. Optimizing the observation policy $\pi$ instead changes the formation process and hence the channel from messages to the acquired record.

\subsection{Phase Transport and Protocol Equivalence}

Phase transport compares response phases at different observation states~\cite{MengPhase2026}. For such a transformation to establish equivalence of communication experiments, it must also be implementable at the receiver and preserve the law of the noisy record. A transformation that depends on an unknown physical state generally fails this requirement.

Suppose two protocols admit transformations of their local records in both directions. Assume that these transformations are independent of unknown messages and environment, are implementable on every time prefix, and preserve admissible policies, record laws, and resource costs. Applying the transformation successively to the observation and decoding operations of any code produces a code for the other protocol with the same message sets, errors, and costs. The reverse construction gives equality of the capacity regions. Each transformation acts only on the corresponding receiver's local record.

The continuous receiver designs below retain a phase reference; the binary model has no continuous phase parameter. Strict separation of its capacity regions rules out such bidirectional implementations between the corresponding policy classes under the same information and budget. A common phase reference preserves consistency of representation, whereas a policy class determines the experiments that can be acquired.

\subsection{Block Channels and Resource Constraints}

In the binary model, the physical and observation states are
\begin{equation}
 X_t=(U_{1,t},U_{2,t},H),\qquad S_t=(S_{1,t},S_{2,t}),
 \qquad\mathcal B=\mathcal X\times\mathcal S.
\end{equation}
The environmental state $H$ is exogenous: a receiver selects its port but cannot change $H$. The encoders $U_i^n=f_i(M_i)$ receive no feedback. Receiver $i$ has the fine record $\mathsf R_i=(V_i,S_i^n,Y_i^n)$ and outputs $\widehat M_i=g_i(\mathsf R_i)$.

Set $z_i=u_i\oplus(h\oplus s_i)u_j$, where $j\ne i$. Conditional on the two-axis state, the receivers acquire independent observations through
\begin{equation}
 K_p(y_1,y_2\mid u_1,u_2,h,s_1,s_2)
 =\prod_{i=1}^2\left[p\ind\{y_i=z_i\}+(1-p)\ind\{y_i=\bot\}\right].
\end{equation}
The retention probability is $p\in[0,1]$; $p=1$ gives the channel without erasures. On finite spaces, the policy-induced channel is
\begin{equation}
\begin{aligned}
&W_n^{f,\pi}(v,s^n,y^n\mid m_1,m_2)\\
&\quad=P_V(v)\sum_{h=0}^1\frac12
 \prod_{t=1}^n\left[
 K_p(y_{1,t},y_{2,t}\mid u_{1,t},u_{2,t},h,s_{1,t},s_{2,t})
 \prod_{i=1}^2\ind\{s_{i,t}=\pi_{i,t}(v_i,s_i^{t-1},y_i^{t-1})\}
 \right].
\end{aligned}
\label{eq:block-law}
\end{equation}
For continuous seeds, the corresponding probability mass is replaced by a measure and integration. All constructions in this paper can use deterministic policies. Equation~\eqref{eq:block-law} retains the action history and the mixture over environmental states.

All schemes in the binary model share the following hard budget: exactly $n$ slots, one binary symbol per transmitter per slot, one port observation per receiver per slot, and at most one port switch per receiver over the block. Denote this budget by $b$. A transmitted zero still occupies a slot; pilots are included in $n$, and an unselected port cannot be read retrospectively. The two transmitters use $2n$ symbol slots and the two receivers make $2n$ observations. These are resource counts rather than energy in joules, since no radio-frequency energy model is specified.

The operational capacity region is the closure of rate pairs achievable by code sequences satisfying this budget, with $R_i=\liminf n^{-1}\log_2N_i$. Maximal joint error takes the maximum over message pairs while averaging over the exogenous state $H$, erasures, and admissible seeds. It is not a worst-case error over $H$. The error threshold $1/2$ below depends on this state-averaged convention.

\section{Bayesian Design of Causal Observations}
\label{sec:policy-optimum}

Instantaneous signal-to-interference-plus-noise ratio describes the power ratio in a single observation. Within a finite communication block, observation design also depends on uncertainty in the responses, the distinguishability of candidate messages, and the remaining switching budget. When the objective is terminal message recovery, the optimization variable on the observation axis is therefore an entire causal policy rather than a single weight vector.

\subsection{Local Posterior and Resource State}

Fix the two users' codebooks, message prior, block length, and common training schedule. Each slot carries a pilot from user 1, a pilot from user 2, or data from both users, as specified before transmission; the receiver's observation choice does not alter the transmitted symbols. Neither transmitter receives feedback, and the receivers do not exchange records. The joint prior of the hidden propagation and array trajectories is assumed known. Receiver thermal noise is independent across slots and independent of the messages, hidden trajectories, and independent receiver random seeds.

For receiver $i$, let $Z_i$ comprise both messages and its local hidden trajectory. The trajectory space is standard Borel, and the responses and noise covariance are measurable functions of the hidden variable. The magnitudes of the individual gains have a common strictly positive lower bound. The complete hidden trajectory is drawn from the given prior, and the receiver updates its knowledge of that trajectory from its own records. The conditional kernel of observation $t$ is
\begin{equation}
 K_{i,t}(\mathrm dy\mid z,w)
 =\mathcal{CN}\!\left(
 w^{\mathrm H}F_{i,t}(z)u_{i,t}(z)
 +w^{\mathrm H}G_{i,t}(z)u_{j,t}(z),
 \sigma^2\|D_{i,t}(z)^{\mathrm H}w\|^2
 \right)(\mathrm dy).
 \label{eq:policy-kernel}
\end{equation}
The conditional mean retains both the desired and interfering symbols; the hidden variable $z$ specifies the codeword structure of both users.

Let $\pi_{i,t}$ be the receiver's posterior on $Z_i$, based on its entire local history before observation $t$. Let $b$ denote the remaining local resources and $v$ the preceding weight vector. Take a nonempty finite set $\mathcal W\subseteq\{w:\|w\|=1\}$, which may represent the amplitude and phase settings available in hardware. Action $w$ consumes $c_t(v,w)$ units of resources, with feasible set
$\mathcal A_t(b,v)=\{w\in\mathcal W:c_t(v,w)\le b\}$; retaining the current weight is assumed always feasible. The numbers of pilots and observations are fixed. If switching is also constrained, its cost component may be set to $\ind\{w\ne v\}$.

Suppress the receiver index. Under posterior $\pi$, the next record has distribution
\begin{equation}
 p_t(\mathrm dy\mid\pi,w)=\int K_t(\mathrm dy\mid z,w)\pi(\mathrm dz).
\end{equation}
Writing $k_t(y\mid z,w)$ for the density in \eqref{eq:policy-kernel}, the posterior update is
\begin{equation}
 \mathcal T_t(\pi,w,y)(\mathrm dz)
 =\frac{k_t(y\mid z,w)\pi(\mathrm dz)}
 {\int k_t(y\mid z',w)\pi(\mathrm dz')}.
 \label{eq:policy-posterior}
\end{equation}
The strictly positive lower bound on the noise variance makes the denominator positive for every finite $y$. Both the unknown responses and the other user's message are integrated against the current joint posterior.

\medskip\noindent\textbf{Proposition (Bayesian observation policy).}
Under the model and prior above, define
\begin{align}
 V_{n+1}(\pi,b,v)&=\max_m\pi\{M_i=m\},\\
 V_t(\pi,b,v)&=\max_{w\in\mathcal A_t(b,v)}
 \int V_{t+1}\bigl(\mathcal T_t(\pi,w,y),
 b-c_t(v,w),w\bigr)\,p_t(\mathrm dy\mid\pi,w).
 \label{eq:policy-bellman}
\end{align}
Then $V_1$ is the largest average probability of correct message recovery attainable with the prescribed weight set and budget. It is attained by a maximizing action in \eqref{eq:policy-bellman} at every step, followed by a maximum a posteriori decision on the desired message.

\begin{proof}
Given the terminal record, maximum a posteriori decoding has conditional success probability $\max_m\pi\{M_i=m\}$. The current posterior retains all information in the past record relevant to the hidden variable and future observations, while $b$ and $v$ determine the remaining feasible actions. For a fixed current action, integration over the next record gives its optimal continuation value. The finite action set ensures that a maximum is attained; a fixed ordering resolves ties measurably. Backward induction then shows that the recursion both bounds the success probability of every causal policy and is attained by the stated policy. Randomized actions yield convex combinations of deterministic action values and cannot improve the optimum.
\end{proof}

The transmitting user's role in each training slot is predetermined, although the receiving direction may depend on past records. The affine predictor introduced below uses fixed training directions. Quantizing its weights to $\mathcal W$ gives a policy covered by the proposition. Experiments with continuous weights use the same information structure, but optimality over a continuous action space requires a separate argument. Full recursion becomes expensive as the hidden trajectory, codebooks, and horizon grow; controllers based on response estimates offer a less costly implementation. Their training error bounds condition on a predetermined training matrix. Adaptive regression must additionally account for dependence between sample selection and noise.

If each receiver has its own budget and neither receiver's weight changes the other receiver's observation kernel, the sum of correctly recovered payloads is separable:
\begin{equation}
 \sup_{\mathsf p_1,\mathsf p_2}
 \E\!\left[k_1\ind\{\widehat M_1=M_1\}
 +k_2\ind\{\widehat M_2=M_2\}\right]
 =k_1V_{1,1}+k_2V_{2,1}.
 \label{eq:policy-separable}
\end{equation}
Equation~\eqref{eq:policy-separable} follows from the product structure of the policy set and the additive objective; the two success events may be correlated. Joint success involves their intersection and generally requires a different optimization. Shared budgets or physical coupling between the observations also destroy this separability. Each receiver's posterior is always formed from its own local records.

\subsection{SINR and Decision Loss}

The two criteria can select different observations even in a single slot. Consider two analog inputs combined into one scalar, with $D=I_2$, $\sigma^2=1$, and known responses
\begin{equation}
 h=(1,2)^{\mathsf T},\qquad g=(0,10)^{\mathsf T},
 \qquad\mathcal W=\{e_1,e_2\}.
\end{equation}
The users independently transmit equiprobable symbols $+1$ and $-1$. Choosing $e_1$ gives $y=u_i+\eta$, with SINR 1; choosing $e_2$ gives $y=2u_i+10u_j+\eta$, with SINR only $4/101$. The instantaneous SINR criterion therefore strictly prefers $e_1$.

For the first observation, the minimum message error probability is
\begin{equation}
 P_{\mathrm e}(e_1)=Q(\sqrt2)\simeq0.07865,
\end{equation}
where $Q(x)=(2\pi)^{-1/2}\int_x^\infty e^{-v^2/2}\,\mathrm dv$.
For the second observation, the four noiseless means are $-12,-8,8,12$. Choosing the nearest mean from the real part and returning its desired-user symbol gives
\begin{equation}
 P_{\mathrm e}(e_2)
 \le Q(2\sqrt2)+\tfrac12Q(8\sqrt2)
 <0.00234.
 \label{eq:policy-counterexample}
\end{equation}
Indeed, conditional on the desired symbol being $+1$, the two means are equiprobably 12 and $-8$. Nearest-mean decoding assigns $+1$ to the real intervals $[-10,0]$ and $[10,\infty)$. Since the real noise variance is $1/2$, the error probability at the first mean is at most $Q(2\sqrt2)$ and at the second at most $Q(2\sqrt2)+Q(8\sqrt2)$. Averaging and applying sign symmetry proves the bound. Optimal decoding of the desired message can only reduce this error probability.

The strong interference in the second observation separates the four transmitted symbol pairs and thereby helps identify the message. Describing interference solely through its power discards this distinguishability. Instantaneous SINR and message decision loss thus define different optimization problems. A finite-horizon observation policy must be chosen jointly with the complete codebook and terminal loss.
\section{Finite-Horizon Observation Policies}
\label{sec:finite-observation-optimum}

An environmental measurement can improve decoding by changing subsequent observations even when it carries no direct information about the message. Its value depends on the remaining horizon and resources. The following finite model permits joint optimization of environmental diagnosis, data observation, and terminal decoding over the full class of causal policies.

\subsection{Diagnostic and Data Observation Model}

Let the message $M$ and environment $H$ be independent and uniform on $\{0,1\}$, with $H$ constant over five slots. The transmitter repeats the same message bit in every slot, has no knowledge of $H$, and receives no feedback. The receiver has two data ports and one local environmental diagnostic port, but can read only one binary output per slot. Write the action as $a\in\{0,1,c\}$. Conditional on $M=m,H=h$, the data ports have kernels
\begin{equation}
 W_a(y\mid m,h)=
 \begin{cases}
  0.9\,\mathbf 1_{\{y=m\}}+0.1\,\mathbf 1_{\{y\ne m\}},&a=h,\\
  1/2,&a\ne h,
 \end{cases}
 \qquad a\in\{0,1\},
\end{equation}
whereas the diagnostic port satisfies
\begin{equation}
 W_c(y\mid m,h)=0.9\,\mathbf 1_{\{y=h\}}
                    +0.1\,\mathbf 1_{\{y\ne h\}}.
\end{equation}
The state $H$ identifies the useful receiving branch: that branch supplies a noisy message bit, while the other contains noise alone. The diagnostic port is an independent sensor of the environment. These three kernels specify the physical observation model. Conditional on $(M,H)$, noise is independent across slots and potential ports; unselected outputs do not enter the receiver's record.

Every action occupies a full slot. Data observations have zero diagnostic cost, while $c$ has unit cost. Every realized record path must satisfy
\begin{equation}
 \sum_{t=1}^{5}\mathbf 1_{\{A_t=c\}}\leq 1.
 \label{eq:finite-diagnostic-budget}
\end{equation}
The controller uses only previous actions and readings, and the terminal decoder uses the same record. All compared schemes share the three ports, five scalar readings, and constraint \eqref{eq:finite-diagnostic-budget}. Open-loop schemes may also perform diagnosis, and their terminal decoders use its result, but subsequent ports follow a predetermined sequence. Port switching consumes no additional time or energy in this example. A switching cost would require an additional resource state and a new optimization.

\subsection{Posterior State and Optimal Recursion}

After any record, the four-component posterior $\pi(m,h)$ describes the receiver's uncertainty about the message and environment. Let $r$ be the number of remaining slots and $b$ the remaining diagnostic budget. If observation terminates now, the optimal success probability is
\begin{equation}
 V_{0,b}(\pi)=\max_{m\in\{0,1\}}\sum_h\pi(m,h).
\end{equation}
Set $\mathcal A_b=\{0,1\}$ for $b=0$, and $\mathcal A_b=\{0,1,c\}$ otherwise. Under action $a$, the next reading has probability
\begin{equation}
 p_a(y\mid\pi)=\sum_{m,h}\pi(m,h)W_a(y\mid m,h),
\end{equation}
and Bayes' rule gives the updated posterior $\pi^{a,y}$. The optimal value of the complete observation process satisfies
\begin{equation}
 V_{r,b}(\pi)=\max_{a\in\mathcal A_b}
 \sum_{y=0}^{1}p_a(y\mid\pi)
 V_{r-1,b-\mathbf 1_{\{a=c\}}}(\pi^{a,y}).
 \label{eq:finite-observation-bellman}
\end{equation}
Equation~\eqref{eq:finite-observation-bellman} values each action by terminal decoding success, accounting for both its immediate reading and the subsequent port choices it permits.

\noindent\textbf{Proposition (Optimal finite-horizon policy).}
Under the preceding model and pathwise budget, \eqref{eq:finite-observation-bellman} equals the largest average success probability attainable by any admissible causal observation policy and terminal decoder. A deterministic policy attains this value.

\noindent\textit{Proof.}
At $r=0$, maximum a posteriori decoding is optimal. Suppose the assertion holds for $r-1$. Given the current record and action, the two possible readings have the stated probabilities, and the induction hypothesis supplies an optimal continuation on each branch. Taking the expected continuation value and maximizing over feasible actions gives \eqref{eq:finite-observation-bellman}. Selecting the corresponding continuation on each branch attains it. Randomized actions give only convex combinations of deterministic values, so deterministic policies suffice. The diagnostic cost is deducted before each continuation, preserving the constraint on every path.

The recursion can be evaluated exactly using unnormalized integer weights. Order the hidden states as $(m,h)=(0,0),(0,1),(1,0),(1,1)$ and start from $q=(1,1,1,1)$. After $t$ readings, the common probability denominator is $4\cdot10^t$. A successor weight is obtained by multiplication by $10W_a(y\mid m,h)\in\{1,5,9\}$. At a terminal node, take the larger of the two message-weight sums; at an intermediate node, sum over the reading branches and maximize over actions. The resulting finite recursion has no rounding error.

\subsection{Optimal Diagnostic Policy}

For five slots, one diagnostic reading, and crossover probability $0.1$, the results are given in Table~\ref{tab:finite-observation-optimum}. The myopic policy maximizes the probability of correct decoding immediately after the next sample, resolving ties in the fixed order $0,1,c$. The open-loop optimum is taken over all $112$ action sequences satisfying the budget, each with its own optimal terminal decoder.

\begin{table}[H]
\centering
\caption{Observation policies in the five-slot diagnostic model.}
\label{tab:finite-observation-optimum}
\begin{tabular}{lrrr}
\toprule
Policy & Average error & Mean diagnoses & Maximum diagnoses\\
\midrule
Fixed data port & $25.428\%$ & $0$ & $0$\\
Myopic & $12.740\%$ & $0.255$ & $1$\\
Optimal open-loop & $10.760\%$ & $1$ & $1$\\
Optimal causal & $7.520\%$ & $1$ & $1$\\
\bottomrule
\end{tabular}
\end{table}

One optimal causal policy first diagnoses the environment, then takes four readings from the indicated data port, and finally applies majority decoding with ties assigned to zero. Diagnosis is correct with probability $0.9$, in which case the four data readings pass through a binary symmetric channel with crossover probability $p=0.1$. When diagnosis is wrong, the selected port contains independent uniform noise and the error probability is $1/2$. The average majority-decoding error for four readings is $3p^2-2p^3$, hence
\begin{equation}
 P_{\mathrm e}^{\star}
 =0.9(3p^2-2p^3)+0.1\cdot\frac12
 =\frac{47}{625}=0.0752.
 \label{eq:finite-optimal-error}
\end{equation}
Conditional on the diagnostic reading, the two message likelihoods contain the same uniform contribution from the uninformative port. It cancels in their comparison, leaving only the informative-port contribution. Majority decoding is therefore the maximum a posteriori rule for this policy.

Global optimality follows from the complete recursion \eqref{eq:finite-observation-bellman}. There are $266$ reachable sufficient states, including $140$ action-decision states. The integer recursion gives root value $369920/(4\cdot10^5)=578/625$, exactly the success probability of the policy in \eqref{eq:finite-optimal-error}. The accompanying data list every state, terminal value, and optimal action; an independent program verifies all successor relations and maximization identities. Exhaustive enumeration of all $3$, $22$, and $1032$ admissible causal trees for horizons one, two, and three gives the same respective optimal values. The five-slot computation uses the sufficient-state reduction, whose completeness follows by backward induction.

The optimal policy is not unique. Constraining the first action to data port $0$, data port $1$, or the diagnostic port and optimizing the remaining actions gives the same root value in all three cases. Diagnosis first is thus a particularly simple realization of an optimal policy.

Relative to the best open-loop sequence, causal observation reduces the error probability from $10.76\%$ to $7.52\%$: an absolute reduction of $3.24$ percentage points and a relative reduction of $81/269\simeq30.11\%$. Both policies use one diagnostic reading and five slots. Their difference is whether the diagnostic result may change how subsequent observations are formed. This comparison uses average error under equal priors; maximum message error with a fixed tie rule is a separate optimization problem. Every scheme transmits one bit in five slots, so the gain is in decoding reliability for the same code and duration.

This finite model shows how environmental identification can reduce terminal message error through subsequent observation choices. Continuous arrays also require continuous posteriors, weight constraints, and evolving environments. To study that extension, we first relate task information to decision loss and then develop finite approximations to continuous actions.
\section{Task Information and Bayesian Observation Design}
\label{sec:task-information-design}

The communication task determines the value of an observation policy. SINR measures separation in a single observation, message mutual information measures the uncertainty removed by the record, and decision risk depends on the decoding loss. These criteria generally select different policies. Under causal observation, environmental diagnosis can reduce terminal risk by changing subsequent record kernels even when the diagnostic record itself is independent of the message.

\subsection{Message Information and Readout Loss}

Let the message $M$ take finitely many values. Fix an implementable policy $\mathsf p$, let $R$ be the complete record actually acquired by the receiver, and let $Z$ be the record supplied to the terminal decoder. If $Z$ is generated from $R$ by a readout kernel independent of the unknown message, the chain rule gives
\begin{equation}
 I_{\mathsf p}(M;Z)
 =I_{\mathsf p}(M;R)-\Delta_M(\mathsf p),
 \qquad \Delta_M(\mathsf p)=I_{\mathsf p}(M;R\mid Z)\ge0.
 \label{eq:task-supply-defect}
\end{equation}
The two terms on the right are the message information supplied by record formation and the loss due to postprocessing, both induced by the same policy. Optimization over the budget-constrained class $\mathfrak P(b)$ therefore takes the form
\begin{equation}
 \sup_{\mathsf p\in\mathfrak P(b)}
 \bigl[I_{\mathsf p}(M;R)-\Delta_M(\mathsf p)\bigr],
 \label{eq:task-joint-optimization}
\end{equation}
The supply and loss terms generally attain their extrema at different policies; they cannot be optimized separately and then subtracted. When $Z=R$, readout loss vanishes, and direction optimization acts directly on the message information produced by record formation.

Here $R$ contains only the records acquired through one sampling chain. The complete analog vector belongs to a different experiment in which all ports are readable. It may supply an information upper bound, but is not part of the implemented controller's information set.

\subsection{Information Decomposition under Causal Observation}

Let $A_t$ be the observation action chosen before reading $t$, let $Y_t$ be the resulting record, and write $\mathcal H_t=(V,A^t,Y^t)$. The initial seed $V$ is known at the receiver and independent of the message. Since the policy depends only on $\mathcal H_{t-1}$,
\begin{equation}
 I(M;A_t\mid\mathcal H_{t-1})=0.
\end{equation}
Successive application of the chain rule yields
\begin{equation}
 I(M;\mathcal H_n)
 =\sum_{t=1}^n I(M;Y_t\mid\mathcal H_{t-1},A_t).
 \label{eq:task-causal-information}
\end{equation}
An observation action affects task information through the conditional kernel of the next record. A diagnostic action with zero immediate information gain may increase the sum of subsequent terms. Maximizing information at each step therefore need not maximize terminal information.

The environmental diagnosis of the preceding section is an example: its record is independent of the message, yet determines the port used for the next four readings. Its benefit appears in the later conditional-information terms of \eqref{eq:task-causal-information}.

\subsection{Logarithmic Loss and Decision Risk}

Let $\pi$ be the joint posterior of message, interference, and environment given the local history, and let $\pi_M$ be its message marginal. Under logarithmic loss, the minimum conditional terminal loss is
\begin{equation}
 h_M(\pi)=-\sum_m\pi_M(m)\log_2\pi_M(m).
\end{equation}
Reporting any distribution $q_M$ incurs conditional loss
$h_M(\pi)+D(\pi_M\Vert q_M)$. Convexity of the negative logarithm makes the divergence nonnegative, with equality at $q_M=\pi_M$. Replacing the terminal value in Section~\ref{sec:policy-optimum} by $h_M(\pi)$ and taking a minimum at every step therefore gives the minimum terminal conditional entropy, or equivalently the maximum task mutual information for the fixed message prior.

Under zero--one loss, terminal risk is
\begin{equation}
 e_M(\pi)=1-\max_m\pi_M(m).
\end{equation}
The functions $e_M(\pi)$ and $h_M(\pi)$ assign different values to posterior uncertainty and generally lead to different optimal policies. Solving the two recursions separately permits a comparison of information and error probability in the same message experiment.

For a binary message, let $E$ be the minimum error probability conditional on the final record, so that $0\le E\le1/2$. The overall minimum error probability is $P_{\rm e}=\E E$, and the conditional entropy is $H_c=\E h_2(E)$. On $[0,1/2]$, binary entropy is concave and satisfies $h_2(e)\ge2e$. Consequently,
\begin{equation}
 h_2^{-1}(H_c)\le P_{\rm e}\le\frac{H_c}{2}.
 \label{eq:task-entropy-error}
\end{equation}
The inverse is taken on $[0,1/2]$. The left inequality follows from $H_c\le h_2(\E E)$ and the right from $H_c\ge2\E E$. These bounds connect task information and decision risk without generally implying that their optimal policies coincide.

A sufficient state for sequential control must also retain the environmental information that determines future record kernels. In the diagnostic model, a known environment of either 0 or 1 leaves the message marginal uniform, yet the optimal data ports are opposite. The control state is thus the joint posterior together with the remaining budget; the message marginal alone is generally sufficient only for the terminal decision.
\section{Finite-Action Approximation of Continuous Observation Policies}
\label{sec:continuous-policy}

Continuous amplitude and phase weights turn finite-action dynamic programming into an approximation problem. A small perturbation of the current action changes the observation law and, through the posterior, affects subsequent observations. The resulting error must therefore be bounded over the entire causal experiment.

Fix the transmit code, the number of readouts, and the resource protocol. The hidden variable $Z$ comprises the desired message, the interfering message, and one of finitely many possible environmental trajectories. Its prior is independent of the receiver's actions. The observation space is standard Borel. Given $Z=z$ and weight $w$, the $t$th observation has measurable kernel $K_{t,w}(\mathrm dy\mid z)$, with conditionally independent noise across readouts. The weight set $\mathcal W$ is compact. The controller chooses actions from its local history to maximize the probability of a correct terminal message decision. The environmental trajectory contained in $Z$ remains hidden.

In the physical--observation dual-axis structure, the transmit code specifies the message path on the physical axis, while the causal weights $w_t$ specify the observation kernels on the observation axis. The optimal value consequently depends on the code, prior, information available to the controller, and cost protocol.

\subsection{Total Variation Bounds for Policy Experiments}

We use the convention
\[
 \|P-Q\|_{\mathrm{TV}}=\sup_A|P(A)-Q(A)|.
\]
Choose a finite set $\mathcal W_\delta\subseteq\mathcal W$ and a fixed measurable rounding map
$q_\delta:\mathcal W\to\mathcal W_\delta$. Assume that rounding preserves the original hard resource constraints along every admissible action path and that
\begin{equation}
 \sup_{z,w}
 \|K_{t,w}(\cdot\mid z)-K_{t,q_\delta(w)}(\cdot\mid z)\|_{\mathrm{TV}}
 \le\omega_t(\delta),\qquad 0\le\omega_t(\delta)\le1.
 \label{eq:continuous-kernel-modulus}
\end{equation}
Let $V^*$ be the optimal probability of a correct decision over all admissible causal policies with continuous weights, and let $V_\delta^*$ be the corresponding value when the weights are restricted to $\mathcal W_\delta$, with the same observations and budget.

\medskip\noindent\textbf{Proposition (Finite-action approximation).}
Under these assumptions,
\begin{equation}
 0\le V^*-V_\delta^*
 \le 1-\prod_{t=1}^n[1-\omega_t(\delta)]
 \le\sum_{t=1}^n\omega_t(\delta).
 \label{eq:continuous-policy-gap}
\end{equation}
In particular, if $\omega_t(\delta)\le\omega(\delta)$, the upper bound becomes
$1-[1-\omega(\delta)]^n\le n\omega(\delta)$. Uniform vanishing of the kernel moduli implies convergence of the finite-action optimal values to the continuous-action value. Attainment of the supremum is not required.

\begin{proof}
Consider any continuous-action policy. Write its actions recursively as
$w_t=f_t(y^{t-1})$. If the original representation also includes past actions, the controller can reconstruct them recursively from the same observation history. Construct an implementable policy that runs this recursion on the observations actually obtained, retains the virtual action $f_t(y^{t-1})$, and applies
$q_\delta(f_t(y^{t-1}))$. This rounded policy is causal and, by assumption, satisfies the original resource constraints.

Couple the two experiments with the same $Z$. As long as their first $t-1$ observations agree, their virtual actions agree. The next observations then follow a pair of kernels covered by~\eqref{eq:continuous-kernel-modulus}. They can therefore be coupled to agree with conditional probability at least $1-\omega_t(\delta)$. Multiplication over the successive readouts bounds the probability of any disagreement by
$1-\prod_t[1-\omega_t(\delta)]$.

The decoder in the rounded experiment can reconstruct the virtual actions from its actual observations and use the original decoding rule. When the coupled observation sequences agree, both the message and the decoding decision agree. The difference in correct-decision probabilities is therefore bounded by the probability of disagreement. Optimizing the decoder for the rounded experiment can only increase its correct-decision probability. Taking the supremum over the original continuous policies gives the upper bound; inclusion of the policy classes gives the lower bound. The final inequality follows by expanding the product or applying a successive union bound. If receiver-side randomization is allowed, couple the same random seed first and apply the same argument.
\end{proof}

When the continuity moduli differ substantially across hidden states, their prior weights yield a sharper bound. If
\[
 \sup_w\|K_{t,w}(\cdot\mid z)-K_{t,q_\delta(w)}(\cdot\mid z)\|_{\mathrm{TV}}
 \le\omega_t(z,\delta),
\]
then
\begin{equation}
 0\le V^*-V_\delta^*
 \le\sum_z\pi_0(z)
 \left\{1-\prod_{t=1}^n[1-\omega_t(z,\delta)]\right\}.
 \label{eq:continuous-prior-weighted-gap}
\end{equation}
To obtain this bound, condition on the common value $Z=z$, couple the complete observation sequences, and average under the original prior. This conditioning is used only to compare probabilities; it does not enlarge the controller's information set.

The distance above is defined on a common observation space, from which the virtual actions are reconstructed recursively. Including distinct deterministic action labels $w$ and $q_\delta(w)$ in the comparison would make the total variation distance between the joint laws equal to~1, obscuring the perturbation of the observation kernels.

A fixed rounding map preserves a hard switching budget:
\begin{equation}
 \sum_{t=1}^n\ind\{q_\delta(w_t)\ne q_\delta(w_{t-1})\}
 \le\sum_{t=1}^n\ind\{w_t\ne w_{t-1}\}.
 \label{eq:continuous-switch-budget}
\end{equation}
If the protocol fixes the initial weight, one must also require $q_\delta(w_0)=w_0$. A fixed cost per readout is likewise preserved. Weight-dependent energy consumption, angular travel, and settling time require separate verification or a budget margin for rounding. Since average costs also depend on the observation law, applying~\eqref{eq:continuous-policy-gap} under an average-cost constraint requires a corresponding cost perturbation bound.

\subsection{Continuity Moduli for Gaussian Observation Kernels}

Consider unit-norm weights and a common noise variance:
\begin{equation}
 K_{t,w}(\cdot\mid z)
 =\mathcal{CN}\bigl(w^{\mathrm H}r_t(z),\sigma^2\bigr),
 \qquad \|w\|=1,\quad \sigma>0.
 \label{eq:continuous-common-noise}
\end{equation}
Here $r_t(z)$ is the noiseless analog response at the $t$th readout, determined by the environment and the complete codeword. If rounding satisfies
$\|w-q_\delta(w)\|\le\delta$, set
$R_t=\max_z\|r_t(z)\|$. Then
\begin{equation}
 \omega_t(\delta)
 \le 2\Phi\!\left(\frac{\delta R_t}{\sqrt2\sigma}\right)-1
 \le \min\left\{1,\frac{\delta R_t}{\sigma\sqrt\pi}\right\},
 \label{eq:continuous-gaussian-modulus}
\end{equation}
where $\Phi$ is the standard real Gaussian distribution function.

\begin{proof}
For the equal-variance complex Gaussian laws $\mathcal{CN}(\mu,\sigma^2)$ and
$\mathcal{CN}(\mu',\sigma^2)$, the region where one density exceeds the other is a half-plane whose boundary passes through the midpoint of the means and is perpendicular to $\mu-\mu'$. The real projection along this direction has variance $\sigma^2/2$. Integration therefore gives the exact distance
\[
 \|\mathcal{CN}(\mu,\sigma^2)-\mathcal{CN}(\mu',\sigma^2)\|_{\mathrm{TV}}
 =2\Phi\!\left(\frac{|\mu-\mu'|}{\sqrt2\sigma}\right)-1.
\]
The means differ by at most $\delta R_t$. Integrating the Gaussian density from~0 and using its upper bound $1/\sqrt{2\pi}$ gives~\eqref{eq:continuous-gaussian-modulus}.
\end{proof}

The bound remains valid if observation formation is followed by a common fixed quantizer or a common randomized readout kernel independent of the unknown parameter: total variation cannot increase under such postprocessing. For a circular phase grid with maximum phase error $\Delta$, the corresponding unit complex numbers differ by a chord of length $2\sin(\Delta/2)$. This chord length is the appropriate value of $\delta$ in the continuity modulus.

If amplitude and phase mismatch change the actual noise variance to
$\sigma^2\|D_t(z)^{\mathrm H}w\|^2$, a change in weight perturbs both the mean and the variance, and both perturbations must enter the distance bound. Reduction to a common-noise section requires known covariance and a normalization compatible with the original hardware and resource constraints. Quotienting out a global phase likewise requires an implementable output phase transformation and a readout protocol that preserves this equivalence.

\subsection{Dynamic Programming and Attainment}

When the protocol supplies a finite observation $a\in\mathcal Y$ and the hidden variable is finite, the finite-grid optimum follows from a finite policy-tree recursion. Represent the joint probability mass of a history branch by the unnormalized vector
$\alpha(z)$, and write
$p_t(a\mid z,w)=K_{t,w}(\{a\}\mid z)$. The terminal value and recursion are
\begin{align}
 F_{n+1}(\alpha,b,v)
 &=\max_m\sum_{z:\,M(z)=m}\alpha(z),\\
 F_t(\alpha,b,v)
 &=\max_{w\in\mathcal A_t(b,v)}
   \sum_{a\in\mathcal Y}
   F_{t+1}\bigl(\alpha\odot p_t(a\mid\cdot,w),
       b-c_t(v,w),w\bigr).
 \label{eq:continuous-unnormalized-dp}
\end{align}
Here $\odot$ denotes componentwise multiplication, and the initial $\alpha$ is the prior of the hidden variable. Branch probabilities are already included in the mass vector, so zero-probability branches contribute zero. Recursion over every feasible action and observation branch yields the global optimum for the specified quantized readout protocol.

For a compact action set without costs that couple successive actions, continuity of each $p_t(a\mid z,w)$ in $w$ also ensures attainment of the continuous-action optimum. The correct-decision probability of any fixed policy, as a function of the initial mass vector, is
$\sum_z\alpha(z)s(z)$ with $0\le s(z)\le1$. Hence the optimal value satisfies
\begin{equation}
 |F_t(\alpha)-F_t(\widetilde\alpha)|
 \le\|\alpha-\widetilde\alpha\|_1.
 \label{eq:continuous-value-lipschitz}
\end{equation}
The terminal value is continuous. Together with the finite observation sum, this bound makes the objective at each backward step jointly continuous in $(\alpha,w)$. Its maximum over the compact action set is therefore attained, and backward induction applies. For the scalar angle used below, take the closed interval $[0,\pi]$ and select the smallest maximizing angle. This selection is measurable: for any $a$, the smallest maximizing angle is at most $a$ if and only if the maximum over $[0,a]$ equals the maximum over the full interval, an equality between continuous functions. The endpoints differ only by a readout sign reversal and hence do not add a distinct physical action. These selections yield a measurable causal policy attaining the continuous optimum.

Suppose the computed grid value is $\widehat V_\delta$ and the numerical error of the recursion has been bounded by
$|\widehat V_\delta-V_\delta^*|\le\varepsilon_{\rm num}$. The continuous-action optimum under the same readout protocol then satisfies
\begin{equation}
 \widehat V_\delta-\varepsilon_{\rm num}
 \le V^*
 \le\min\left\{1,\widehat V_\delta+\varepsilon_{\rm num}
       +1-\prod_t[1-\omega_t(\delta)]\right\}.
 \label{eq:continuous-numerical-certificate}
\end{equation}
Equation~\eqref{eq:continuous-numerical-certificate} supplies a numerical stopping criterion. The policy tree still grows with the number of readouts.

\subsection{A Binary-Quantized Gaussian Observation Model}

Let $Z=(M,J,H)$ remain fixed within the block, with independent uniform variables $M,J\in\{-1,+1\}$ and $H\in\{0,1\}$. At each of the three readouts, the physical vector is
\begin{equation}
 r_t=Mh_H+Jg_H+\nu_t,\qquad
 \nu_t\sim\mathcal N(0,I_2),
 \label{eq:continuous-real-example}
\end{equation}
with independent noise across readouts. Set
\begin{align*}
 h_0&=(1.4,0.5)^{\mathsf T},&g_0&=(0.3,2.2)^{\mathsf T},\\
 h_1&=(0.5,1.4)^{\mathsf T},&g_1&=(2.2,0.3)^{\mathsf T}.
\end{align*}
The receiver observes only $Y_t=\operatorname{sign}(w_{\theta_t}^{\mathsf T}r_t)$, where
$w_\theta=(\cos\theta,\sin\theta)^{\mathsf T}$. For $a\in\{-1,+1\}$, the likelihood is
\begin{equation}
 p(a\mid m,j,h,\theta)
 =\Phi\!\left[a\,w_\theta^{\mathsf T}(mh_h+jg_h)\right].
 \label{eq:continuous-sign-likelihood}
\end{equation}
The terminal decision in~\eqref{eq:continuous-unnormalized-dp} sums over both $J$ and $H$ to recover the desired message $M$. Neither the propagation state nor the interfering symbol is available to the controller.

Since $w_{\theta+\pi}=-w_\theta$, adding $\pi$ to the angle merely reverses the comparator output. The controller knows its own action and can implement this reversal. The $K$ equally spaced directions
$\theta_k=k\pi/K$ have covering radius, modulo this equivalence,
\begin{equation}
 \delta_K=2\sin\!\left(\frac{\pi}{4K}\right),
 \qquad
 R=\max_{m,j,h}\|mh_h+jg_h\|=\sqrt{10.18}.
 \label{eq:continuous-projective-radius}
\end{equation}
Rounding across the boundary between $0$ and $\pi$ includes a readout sign reversal. If the nearest vector is
$\epsilon w_{\theta_k}$, with $\epsilon\in\{-1,+1\}$, the receiver applies
$w_{\theta_k}$ and multiplies its observation by the known sign $\epsilon$. The resulting virtual observation is passed to the original policy recursion. Both experiments are thereby expressed in the same virtual observation coordinates.

For equal-variance real Gaussian laws,
\[
 \|\mathcal N(\mu,1)-\mathcal N(\mu',1)\|_{\mathrm{TV}}
 =2\Phi(|\mu-\mu'|/2)-1.
\]
Define
\begin{equation}
 \omega_K=2\Phi(\delta_K R/2)-1,
 \qquad G_K=1-(1-\omega_K)^3.
 \label{eq:continuous-real-gap}
\end{equation}
Among the eight hidden states, half have noiseless response norm $R_+=\sqrt{10.18}$ and half have norm $R_-=\sqrt{4.10}$. Equation~\eqref{eq:continuous-prior-weighted-gap} therefore gives the sharper bound
\begin{equation}
 \overline G_K=\frac12\sum_{s\in\{+,-\}}
 \left[1-\left\{2-2\Phi(\delta_K R_s/2)\right\}^3\right]
 \le G_K.
 \label{eq:continuous-real-weighted-gap}
\end{equation}
If $e_K$ is the minimum average message error probability over all three-step causal policies using the $K$ directions, the optimum $e_{\rm cont}$ over continuous directions with one-bit readouts satisfies
\begin{equation}
 \max\{0,e_K-\overline G_K\}\le e_{\rm cont}\le e_K.
 \label{eq:continuous-real-error-interval}
\end{equation}
A bounded numerical error in the recursion is included by subtracting it from the lower endpoint and adding it to the upper endpoint. Both sides refer to the same three-slot, one-bit readout protocol; refinement of the action grid narrows the interval.

The same rounding preserves a predetermined observation sequence, so the bound also applies to the open-loop optimum. Writing
$e_{\rm open,K}$ and $e_{\rm causal,K}$ for the error probabilities of the best predetermined sequence and the best causal policy on the grid gives
\begin{equation}
 e_{\rm open,cont}-e_{\rm causal,cont}
 \ge e_{\rm open,K}-e_{\rm causal,K}-\overline G_K.
 \label{eq:continuous-strict-adaptive-gain}
\end{equation}
A positive right-hand side establishes a strict separation between causal policies and all continuous open-loop sequences.

\subsection{Kernel Interpolation and Second-Order Approximation}

With a fixed number of readouts and no additional switching constraint or weight-dependent cost, randomization between adjacent grid actions cancels the first-order kernel approximation error. The policy approximation error then decreases as $K^{-2}$.

Let $h=\pi/K$. If the virtual continuous policy chooses
$\theta\in[\theta_k,\theta_{k+1}]$ at the current history, apply the two endpoint directions with probabilities
\begin{equation}
 \lambda=\frac{\theta_{k+1}-\theta}{h},\qquad 1-\lambda,
 \label{eq:continuous-mixture-weights}
\end{equation}
respectively. Each fresh random draw is independent of the hidden state, past history, and receiver noise; its selection probabilities depend on the current history. The controller passes the resulting bit to the original continuous-policy recursion without additionally using the chosen endpoint index in the virtual posterior. Conditional on $Z=z$ and the current virtual observation history, the probability of a $+1$ output is exactly
\begin{equation}
 \widetilde p_\theta(z)
 =\lambda p_{\theta_k}(z)+(1-\lambda)p_{\theta_{k+1}}(z),
 \qquad p_\theta(z)=\Phi(\mu_\theta(z)),
 \label{eq:continuous-mixture-kernel}
\end{equation}
where $\mu_\theta(z)=w_\theta^{\mathsf T}(mh_H+jg_H)$.
The right endpoint of the final grid interval is $\pi$. If it is selected, the receiver uses direction~0 and reverses the comparator output. This reversal depends only on the controller's action, not on the unknown state.

\medskip\noindent\textbf{Proposition (Second-order policy approximation).}
For the one-bit readout protocol in~\eqref{eq:continuous-real-example}, define
\begin{align}
 A_R&=1+R^2,\\
 u_R&=\min\left\{R,
 \sqrt{\frac{A_R+3-\sqrt{(A_R+3)^2-4A_R}}{2}}\right\},\\
 L(R)&=\frac{e^{-u_R^2/2}}{\sqrt{2\pi}}
       u_R(1+R^2-u_R^2),\\
 \varepsilon_K(R)&=\min\left\{1,\frac{\pi^2L(R)}{8K^2}\right\}.
 \label{eq:continuous-second-order-modulus}
\end{align}
With $R(z)=\|mh_H+jg_H\|$,
\begin{equation}
 0\le e_K-e_{\rm cont}
 \le G_K^{(2)}
 :=\sum_z\pi_0(z)
       \left[1-\{1-\varepsilon_K(R(z))\}^3\right].
 \label{eq:continuous-second-order-gap}
\end{equation}
In this example, the right-hand side is the equally weighted average of the expressions for $R_+$ and $R_-$. Here $e_K$ remains the optimal error probability over \emph{deterministic} causal grid policies; the bound also holds for the optimal values restricted to deterministic policies.

\begin{proof}
For $f\in C^2([a,b])$, let $\ell$ be its linear interpolant through the endpoints. At a given interior point $x$, subtract from $f-\ell$ the quadratic that vanishes at $a,b$ and agrees with $f-\ell$ at $x$. The resulting function has three zeros. Two applications of Rolle's theorem give some $\xi\in(a,b)$ such that
\[
 f(x)-\ell(x)=\tfrac12f''(\xi)(x-a)(x-b).
\]
Consequently, $|f-\ell|\le (b-a)^2\sup|f''|/8$.

Fix $z$, write $\mu_\theta=R\cos(\theta-\alpha)$, and let
$\phi(u)=e^{-u^2/2}/\sqrt{2\pi}$. Differentiation gives
\begin{equation}
 p''_\theta
 =-\phi(\mu_\theta)\mu_\theta
     \left[1+(\mu'_\theta)^2\right],
 \qquad
 \sup_\theta|p''_\theta|
 =\max_{0\le u\le R}\phi(u)u(1+R^2-u^2).
 \label{eq:continuous-likelihood-curvature}
\end{equation}
The derivative of the expression on the right has the sign of
$u^4-(A_R+3)u^2+A_R$. The larger root in $u^2$ exceeds $R^2$, so the maximum over the interval occurs at the smaller positive root in $u$ or at the right endpoint, whichever comes first. This is $u_R$. Endpoint interpolation therefore satisfies
$|p_\theta(z)-\widetilde p_\theta(z)|\le\varepsilon_K(R(z))$.
For binary distributions, total variation is the absolute difference of either component probability. The same quantity thus bounds the one-step observation-kernel distance.

Couple the virtual observations of the continuous policy and the randomized grid policy for each hidden state, then average over the prior. This gives the upper bound in~\eqref{eq:continuous-second-order-gap}. The construction preserves the number of readouts and the transmitted symbols. At each step, the finite-action Bellman recursion maximizes the action value, whereas local randomization takes a convex combination of those values. Equivalently, draw independent seeds in advance for every node of the finite-depth tree. Fixing all seeds produces an admissible deterministic grid tree. The correct-decision probability of the randomized policy is the average over these deterministic trees, and cannot exceed the deterministic grid optimum. Optimizing over the original continuous policy proves the claim.
\end{proof}

Adjacent-action interpolation also applies to predetermined observation sequences. Independent endpoint draws made in advance define a probability mixture of predetermined grid sequences, at least one of which attains the mixture's average correct-decision probability. Hence
$\overline G_K$ in~\eqref{eq:continuous-strict-adaptive-gain} may be replaced by the smaller $G_K^{(2)}$.
Numerically, $L(R_+)\simeq2.4840893965$ and $L(R_-)\simeq1.0350050969$. For
$K=128,512,1024$, the second-order bounds are
\begin{equation}
 G_{128}^{(2)}\simeq3.9741541\times10^{-4},\quad
 G_{512}^{(2)}\simeq2.4842072\times10^{-5},\quad
 G_{1024}^{(2)}\simeq6.2105631\times10^{-6}.
 \label{eq:continuous-second-order-numbers}
\end{equation}
The last value corresponds to $0.0006211$ percentage points of error probability. It bounds the discretization error for three-step continuous policies; numerical arithmetic errors must be accounted for separately.

Random endpoint selection can increase the number of switches. Under a hard switching budget, weight-dependent energy cost, or settling-time constraint, the second-order bound therefore also requires feasibility of the interpolated policy. Deterministic rounding, by contrast, satisfies~\eqref{eq:continuous-switch-budget}. The three-slot protocol considered here permits at most two inter-slot switches for every policy.

These approximations keep the readout protocol fixed. One-bit and full analog readouts define different experiments, and the gap between their optimal values requires a separate bound. For a fixed fine experiment and its final coarsening $Q$, the chain rule gives
\[
 I(M;R)=I(M;Q)+I(M;R\mid Q),
\]
but this information loss is not the difference between the minimum error probabilities. Different causal readout protocols also induce different subsequent actions. Comparing their optimal values requires a sequentially implementable, parameter-independent observation recovery kernel, together with bounds on its recovery error and resource consumption.
\subsection{Policy-Vector Envelopes and Second-Order Bounds}
\label{sec:alpha-vector-refinement}

For finite observations, let $v(z)$ denote the conditional probability of a correct decision under a given policy tree when the hidden state is $z$. The vector satisfies $0\le v(z)\le1$, and its correct-decision probability under an unnormalized mass vector $\alpha$ is $\langle\alpha,v\rangle$.

Suppose $k$ binary readouts remain, with no switching, angular-travel, or other constraints coupling the actions. Write $p_\theta(z)=\Pr\{Y=+1\mid z,\theta\}$. The set of all deterministic policy vectors is given recursively by
\begin{align}
 \Gamma_0&=\{(\ind\{M(z)=m\})_z:m\in\mathcal M\},\\
 \Gamma_k&=\left\{p_\theta\odot v_++(1-p_\theta)\odot v_-:
 \theta\in\Theta,\quad v_+,v_-\in\Gamma_{k-1}\right\},
 \label{eq:alpha-policy-recursion}
\end{align}
and $F_k(\alpha)=\max_{v\in\Gamma_k}\langle\alpha,v\rangle$. With finitely many actions, this is the upper envelope of finitely many linear functions. For a compact continuous action set and continuous formation kernels, $\Gamma_k$ remains compact, although its envelope need not have finitely many pieces.

The finite-action recursion admits the incremental pruning method of Cassandra, Littman, and Zhang~\cite{cassandra1997}. Vector backups, linear-programming dominance tests, and branchwise pruning preserve the upper envelope of the value function; see Section~2, Eqs.~(6)--(8), Section~3, Eq.~(9) and Fig.~2, and Section~4, Eq.~(10), of that paper. The present model has hidden state $Z$, grid directions as actions, identity state transitions, and rewards concentrated at the terminal message decision. Removing vectors that never raise the upper envelope preserves the optimal correct-decision probability for every prior. This can reduce the computation, although combinatorial growth remains possible in the worst case. The method applies to finite action sets; approximation of continuous angles additionally requires regularity of the observation kernel.

The policy-vector representation also yields a second-order bound by deterministic action rounding. Let $\Theta=[0,\pi]$, let the grid include both endpoints with maximum spacing $h$, and assume
\[
 p_\theta(z)\in C^2,\qquad
 \sup_{\theta\in\Theta}|p_\theta''(z)|\le L_z.
\]
For sign readouts, the endpoint $\pi$ is implemented by direction~0 followed by a known readout sign reversal. If $F_{k,h}$ denotes the grid optimum, then
\begin{equation}
 0\le F_k(\alpha)-F_{k,h}(\alpha)
 \le\frac{k h^2}{8}\sum_z\alpha(z)L_z.
 \label{eq:alpha-deterministic-quadratic}
\end{equation}
Only twice differentiability of the observation kernel is required; the optimal-value envelope need not be differentiable.

\begin{proof}
Take a current continuous optimal angle $\theta_*$ and fix the optimal continuation vectors $v_+,v_-$ on its two observation branches. Define
\[
 f(\theta)=\sum_z\alpha(z)
 \{p_\theta(z)v_+(z)+[1-p_\theta(z)]v_-(z)\}.
\]
Both subtrees remain feasible when the current angle changes. Hence
$f(\theta)\le F_k(\alpha)=f(\theta_*)$ for every $\theta$. If $\theta_*$ is interior, the smooth function $f$ attains its maximum there and satisfies $f'(\theta_*)=0$. If the optimum is at an endpoint, that point already belongs to the grid. Since $|v_+(z)-v_-(z)|\le1$,
\[
 |f''(\theta)|\le\sum_z\alpha(z)L_z.
\]
Round an interior optimal angle to its nearest grid point $\widehat\theta$. Taylor's formula gives
\[
 f(\theta_*)-f(\widehat\theta)
 \le\frac{h^2}{8}\sum_z\alpha(z)L_z.
\]
Now use an optimal grid subtree on each of the two actual posterior branches. By the induction hypothesis, their combined loss is at most
\[
 \frac{(k-1)h^2}{8}\sum_z\alpha(z)L_z
 \{p_{\widehat\theta}(z)+1-p_{\widehat\theta}(z)\}
 =\frac{(k-1)h^2}{8}\sum_z\alpha(z)L_z.
\]
Adding the current-step loss proves the bound. The two values agree when $k=0$. Every choice is deterministic, so a deterministic grid policy attaining the stated bound exists.
\end{proof}

The second-order accuracy comes from the smooth supporting function associated with the optimal subtrees, regardless of whether the full optimal-value envelope is differentiable. Equation~\eqref{eq:alpha-deterministic-quadratic} compares the continuous and discrete optima; implementation requires only finite-grid recursion. The essential condition is that the fixed continuation subtrees remain feasible after perturbing the current action. If future action sets depend on the remaining switching budget or settling time, those resources must be included in the state and feasibility established again.
\section{Numerical Optimization of Continuous Observation Directions}
\label{sec:continuous-experiment}

Consider the three-reading model in \eqref{eq:continuous-real-example}. Each sender repeats one equiprobable bit three times, and the receiver recovers only $M$. The propagation branch $H$ and interfering bit $J$ remain fixed within the block and are initially unknown to the receiver. Every slot uses a real unit-norm combining direction, one sampling chain, and a zero-threshold comparator; three slots yield exactly three bits. The transmitters receive no feedback, and the receiver selects each new direction using only its preceding bits. State identification and message decoding share these three readings. There are no additional pilots, and switching directions incurs no separate cost.

With the transmitted code and block length fixed, the physical superposition law is unchanged. The freedom on the observation axis lies in the dependence of the combining direction on the record history. The design variable is the complete causal direction tree.

\subsection{Policy Space and Optimal Direction Tree}

On uniform grids of $K=32,64,128,256,512$ directions, we separately minimize terminal message error and maximize terminal task information. Causal policies are computed by full backward recursion, and open-loop policies by exhaustive enumeration of predetermined direction triples. Conditional on the hidden state, this model is memoryless and stationary, and it has no order-dependent costs. Permuting an open-loop triple therefore leaves the terminal experiment unchanged, so only nondecreasing triples need be enumerated. A fixed-direction policy uses the same direction in all three slots. Two further baselines greedily minimize the next-step message error or maximize the immediate information gain. For these baselines, ties within tolerance $10^{-12}$ are resolved by choosing the smallest angle. This rule applies only to the greedy policies; the full recursion takes the actual numerical maximum.

Error minimization is continued to $K=1024$, while task-information optimization stops at 512 points. The optimal causal direction tree on the 1024-point grid is
\begin{align}
 \theta_1&=86.66015625^\circ,\\
 \theta_2&=26.19140625^\circ,\\
 \theta_3&=\begin{cases}
 5.625^\circ,&Y_1=Y_2,\\
 38.49609375^\circ,&Y_1\ne Y_2.
 \end{cases}
 \label{eq:continuous-computed-policy}
\end{align}
The terminal decoder marginalizes the joint probabilities of the eight states $(M,J,H)$ and makes a maximum a posteriori decision on $M$. The first two directions are independent of the record; the third depends on whether the first two bits agree. This structure emerges from optimization over the full policy class.

\subsection{Continuous Optima and Separation from Open-Loop Observation}

On the 1024-point grid, the optimal causal error probability is $0.205728260154$, the optimal open-loop error probability is $0.216937702161$, and the best fixed-direction error probability is $0.260267090357$. The causal policy reduces error by approximately $5.1671\%$ relative to the best open-loop policy on the same grid, and by approximately $20.9549\%$ relative to the best fixed direction. Both relative improvements refer to the common finite grid.

The second-order bound $G_{1024}^{(2)}\simeq6.2105631\times10^{-6}$ limits the possible improvement on passing from the grid to continuous directions. Combining it with the computed grid optima gives the interval estimates in Table~\ref{tab:continuous-interval}. The displayed endpoints are rounded outward.

\begin{table}[H]
\centering\small
\caption{Interval estimates for optimal continuous-direction error probabilities (\%), including the analytic discretization bound.}
\label{tab:continuous-interval}
\begin{tabular}{lrr}
\toprule
Admissible policy class & Lower endpoint & Upper endpoint\\
\midrule
All continuous causal policies & 20.57220495 & 20.57282602\\
All continuous predetermined sequences & 21.69314915 & 21.69377022\\
\bottomrule
\end{tabular}
\end{table}

For the exact grid optima, the same bound gives
\begin{equation}
 e_{\rm open,cont}-e_{\rm causal,cont}
 \ge e_{\rm open,1024}-e_{\rm causal,1024}-G_{1024}^{(2)}
 \simeq0.01120323144.
 \label{eq:continuous-final-separation}
\end{equation}
The computed separation is approximately $1.1203$ percentage points, far larger than the discretization bound of $0.0006211$ percentage points. The comparison extends to all continuous open-loop sequences.

\begin{figure}[H]
\centering\includegraphics[width=.93\textwidth]{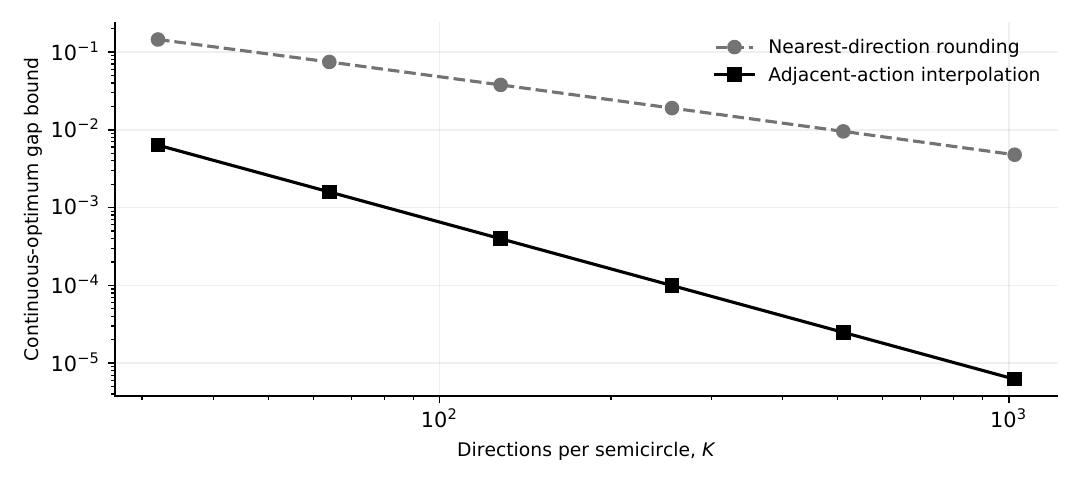}
\caption{Action-discretization bounds for continuous policies. Solid line: second-order interpolation between adjacent actions. Dashed line: prior-weighted nearest-direction rounding.}
\label{fig:continuous-optimality}
\end{figure}

\subsection{Decision Risk, Task Information, and Resources}

\begin{table}[H]
\centering\small\setlength{\tabcolsep}{3.5pt}
\caption{Decision performance and resource use on a 512-point direction grid. Goodput is measured in bit/use and task mutual information in bits. Every policy uses three reads and one sampling chain; switch counts are expectations.}
\label{tab:continuous-metrics}
\begin{tabular}{lrrrr}
\toprule
Policy & Error (\%) & \shortstack{Task mutual\\information} & Goodput & Switches\\
\midrule
Best fixed direction & 26.0267 & 0.184150 & 0.246578 & 0\\
Open-loop error optimum & 21.6939 & 0.317859 & 0.261020 & 2\\
Open-loop information optimum & 25.7467 & 0.348210 & 0.247511 & 1\\
Greedy error reduction & 25.3721 & 0.338652 & 0.248760 & 1\\
Greedy information gain & 22.6198 & 0.359022 & 0.257934 & 2\\
Causal error optimum & 20.5728 & 0.338712 & 0.264757 & 2\\
Causal information optimum & 22.0516 & 0.361707 & 0.259828 & 2\\
\bottomrule
\end{tabular}
\end{table}

Table~\ref{tab:continuous-metrics} compares the two objectives on the same 512-point grid. The information-maximizing policy yields $0.361707177$ bit, whereas the error-minimizing policy yields approximately $0.338711654$ bit. The former nevertheless has the higher error probability: approximately $22.0516\%$, against $20.5728\%$. Thus the two objectives select different policies, as they also do in the open-loop class. The terminal loss should reflect the communication task.

For every computed policy, the conditional error probabilities of the two messages agree to numerical precision, so its average and maximum message error probabilities coincide. Global optimization still uses average error under the prescribed uniform prior. Every scheme uses three scalar readings, one sampling chain, and three slots; each sender transmits one real symbol per slot, and the desired user's nominal rate is $1/3$ bit/slot. Goodput is defined as $(1-P_{\rm e})/3$ and scored against the true message, without acknowledgments or retransmission. Energy efficiency in joules would require a physical energy model for the combining network, comparator, and switching operations.

The metrics reveal several tradeoffs. A fixed direction requires no switching, while the optimal causal policy switches twice. The error-minimizing policy conveys less mutual information, and greedy policies reduce computation at a performance cost. The loss function and resource constraints determine which tradeoff is appropriate.

\subsection{Numerical Accuracy and Scope}

All performance measures in this section are finite probability sums, with no Monte Carlo sampling error. The program uses unnormalized posteriors and complete recursion over feasible actions and record branches. An independent program enumerates all 16384 three-step direction trees and 64 ordered open-loop sequences on a four-point grid. Both objectives agree with the recursion, with maximum discrepancy approximately $1.1\times10^{-16}$. A separate calculation using 90-digit decimal arithmetic independently reconstructs the 12288 Gaussian probabilities on the 512- and 1024-point grids; the maximum discrepancy is below $5.0\times10^{-16}$. Full leaf-by-leaf evaluation of the saved policies gives error-probability discrepancies below $5.3\times10^{-17}$. Additional interpolation checks verify curvature, output-bit reversal across $\pi$, and implementability after the random seed is fixed.

Equation~\eqref{eq:continuous-second-order-gap} is an analytic bound; the tabulated intervals and the separation of $1.1203$ percentage points use floating-point recursion values. High-precision cross-checks do not constitute interval-arithmetic certification of the complete computation. Rigorous numerical endpoints would additionally require bounds on floating-point errors in both recursions and directed rounding of the discretization bound.

The policies and discretization bounds apply to continuous directions with a fixed repetition code, a finite hidden state, and three one-bit readings. Operational capacity with arbitrary transmitted codes and block lengths requires a further optimization, together with converse and achievability arguments.
\section{Linear Observation-Channel Design}
\label{sec:front-design}

Port selection presupposes a family of observation kernels. Receiver design also allows this family to be shaped. Once the combining weights, training schedule, and readout rule are included in the observation state, optimization along the observation axis becomes a choice among receiver channels. We study this choice under a single sampling chain constraint, first with known responses and then with response estimation and its training cost included in the finite-block protocol.

\subsection{Single Sampling Chain Model}

Each receiver has $L$ analog inputs. The physical field in the current slot is
\begin{equation}
 r_i=h_i u_i+g_i u_j+\nu_i,\qquad
 \nu_i\sim\mathcal{CN}(0,\sigma^2I_L),\quad j\ne i.
 \label{eq:front-field}
\end{equation}
The vectors $h_i$ and $g_i\in\mathbb C^L$ are the spatial responses of the desired and interfering users, respectively; they include the propagation gains and the square roots of the fixed transmit powers. The symbols $u_i$ have unit average power. The transmitters have independent messages, share no messages, and receive no feedback; the receivers do not exchange records. The spatial responses remain fixed within a block. Thermal noise is independent across slots and receivers, and independent of the messages and spatial responses. Training and data transmission obey the same noise model.

In the dual-axis representation, $(u_1,u_2,h_i,g_i)$ belongs to the physical state, whereas the complex weights $w_i$, training order, and readout rule belong to the observation state. Amplitude and phase weighting of the four analog inputs produces the scalar record
\begin{equation}
 y_i=w_i^{\mathrm H}r_i
 =a_i(w_i)u_i+b_i(w_i)u_j+\eta_i,
 \quad a_i(w)=w^{\mathrm H}h_i,\quad b_i(w)=w^{\mathrm H}g_i.
 \label{eq:front-scalar}
\end{equation}
The constraint $\|w_i\|_2=1$ gives $\eta_i\sim\mathcal{CN}(0,\sigma^2)$, so that the noise variance is the same for every choice of weights. Only this single complex record is digitized in each slot. Training and data reception use the same scalar interface; the four analog inputs are not separately available as digital records.

The formation kernel is therefore
\begin{equation}
 K_w(\mathrm dy\mid u_i,u_j,h_i,g_i)
 =\mathcal{CN}\bigl(w^{\mathrm H}h_i u_i+w^{\mathrm H}g_i u_j,\sigma^2\bigr).
\end{equation}
The resource constraint is essential to the meaning of observation optimization. If the full vector $r_i$ has already been digitized, linear combining is merely post-processing of the complete record. With a single sampled output, however, $w$ determines which statistical experiment is actually observed. It changes the relative responses of the desired and interfering users before sampling. An invertible rescaling of $y$ after sampling scales the signal, interference, and noise together and leaves the signal-to-interference-plus-noise ratio unchanged.

\begin{figure}[htbp]
\centering\includegraphics[width=.97\textwidth]{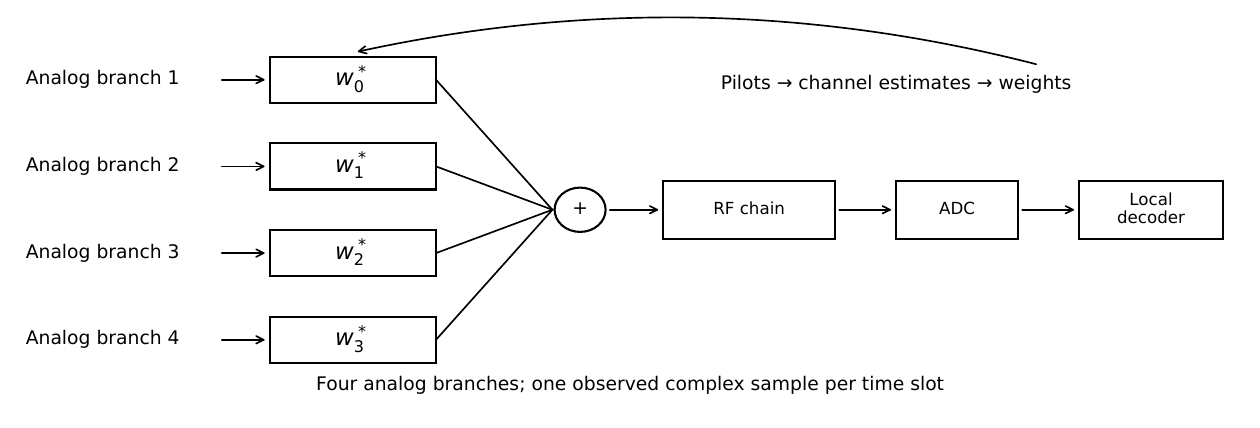}
\caption{Observation front end with a single sampling chain. Complex weights combine four analog inputs; scalar pilot observations provide response estimates from which the data weights are chosen. The diagram represents the signal model and omits RF component nonidealities.}
\label{fig:front-hardware}
\end{figure}

\subsection{Optimal Weights and Spatial Separability}

Suppose first that the true spatial responses are known, and omit the user index. Define
\begin{equation}
 R=gg^{\mathrm H}+\sigma^2I_L,\qquad
 \gamma(w)=\frac{|w^{\mathrm H}h|^2}{w^{\mathrm H}Rw}.
 \label{eq:front-sinr}
\end{equation}
The quantity $\gamma$ is the ratio of desired-signal power to interference-plus-noise power in the record; its definition does not require Gaussian inputs. If the two inputs are additionally independent, circularly symmetric complex Gaussian variables with unit variance, then, conditional on the true $h,g,w$,
$I(u_i;y_i\mid h,g,w)=\log_2(1+\gamma(w))$. This follows from the scalar complex Gaussian entropy formula. The experiments below use finite codes, whose rates must account for training overhead. Gaussian mutual information serves only as a measure of observation quality.

\medskip\noindent\textbf{Proposition (Optimal linear observation for known responses).}
If $\sigma^2>0$ and $h\ne0$, then over all unit-norm complex weights,
\begin{equation}
 w_* =\frac{R^{-1}h}{\|R^{-1}h\|_2},\qquad
 \max_{\|w\|=1}\gamma(w)=h^{\mathrm H}R^{-1}h.
 \label{eq:front-optimum}
\end{equation}
\begin{proof}
Set $v=R^{1/2}w$ and $q=R^{-1/2}h$. Then
$|w^{\mathrm H}h|^2=|v^{\mathrm H}q|^2\le\|v\|^2\|q\|^2$; division by
$w^{\mathrm H}Rw=\|v\|^2$ gives the upper bound. Equality holds if and only if $v$ and $q$ are collinear. Normalization gives~\eqref{eq:front-optimum}. With a single interferer, the expression reduces to $w_*\propto h-g(g^{\mathrm H}h)/(\sigma^2+\|g\|^2)$, without requiring a general matrix inversion.
\end{proof}

The existence of an interference-free observation depends on the two spatial responses. For $g\ne0$, let
$\Pi_g^\perp=I-gg^{\mathrm H}/\|g\|^2$. There exists a weight satisfying both
$w^{\mathrm H}g=0$ and $w^{\mathrm H}h\ne0$ if and only if
$\Pi_g^\perp h\ne0$, or equivalently, $h$ does not lie in the span of $g$. Necessity follows from orthogonality, and sufficiency follows from the construction
\begin{equation}
 w_{\rm null}=\frac{\Pi_g^\perp h}{\|\Pi_g^\perp h\|},\qquad
 \gamma_{\rm null}=\frac{\|h\|^2(1-\rho^2)}{\sigma^2},\qquad
 \rho^2=\frac{|g^{\mathrm H}h|^2}{\|g\|^2\|h\|^2}.
 \label{eq:front-null}
\end{equation}
The factor $1-\rho^2$ measures the signal power retained after interference nulling. As the responses approach collinearity, the retained desired-signal component vanishes. At exact collinearity, nulling the interference necessarily nulls the desired signal as well, and no linear observation can separate them. If $g=0$, take $w=h/\|h\|$.

The weights in~\eqref{eq:front-optimum} balance interference suppression against signal preservation; hard nulling instead requires an exactly zero interference coefficient. Under strong interference the two choices may be very close. Their performance difference measures the gain that SINR optimization provides beyond spatial separation itself.

\subsection{Orthogonal Pilots and Response Estimation}

Let $q_0,\ldots,q_{L-1}$ be unit-norm training weights, each used for $r$ consecutive slots. User~1 first transmits a known symbol while user~2 transmits zero; the roles are then reversed. Each receiver thus estimates its local $h$ and $g$ separately. The total training length is $m=2Lr$, with only one scalar record per slot.

Take the $\ell$th row of the training matrix $A$ to be $q_\ell^{\mathrm H}$. The averaged observations satisfy
$\bar z=Ah+e$, where $e\sim\mathcal{CN}(0,\sigma^2I_L/r)$. If $A$ is invertible, the error covariance of the linear unbiased estimate is
\begin{equation}
 \operatorname{Cov}(\widehat h-h)
 =\frac{\sigma^2}{r}(A^{\mathrm H}A)^{-1}.
\end{equation}
Unit-norm rows imply $\operatorname{tr}(A^{\mathrm H}A)=L$. Its positive eigenvalues $\lambda_1,\ldots,\lambda_L$ obey
$(\sum\lambda_\ell)(\sum\lambda_\ell^{-1})\ge L^2$, whence
\begin{equation}
 \E\|\widehat h-h\|^2\ge\frac{L\sigma^2}{r}.
\end{equation}
Equality requires every eigenvalue to equal one, which is precisely orthogonality of the training observations. Orthogonal training therefore minimizes total mean-square error among linear unbiased designs without a prior. Choosing
\begin{equation}
 (q_\ell)_a=L^{-1/2}\exp(2\pi\mathrm i\ell a/L),
 \qquad a,\ell=0,\ldots,L-1,
 \label{eq:front-probes}
\end{equation}
yields $\widehat h=\sum_\ell q_\ell\bar z_\ell$; the same construction estimates $g$. For $L=4$, training requires only four equal-amplitude phase settings.

The receiver estimates $\widehat h,\widehat g$ during the first $2Lr$ slots and then chooses the data weights. In this causal schedule, pilots serve response estimation alone and carry no user message. Grouping repetitions of each weight produces $2L$ consecutive training segments over the two rounds, followed by one data segment. Each receiver changes its weights at most $2L$ times. The binary model introduced later has a separate budget of one port switch; these resource constraints belong to different models.

\subsection{Robust Observation under Bounded Estimation Error}

Substituting the estimated responses into~\eqref{eq:front-optimum} maximizes nominal SINR. To account for response uncertainty, consider the error sets
\begin{equation}
 \|h-\widehat h\|\le\epsilon_h,\qquad
 \|g-\widehat g\|\le\epsilon_g.
\end{equation}
For any unit-norm weight, the exact worst-case SINR is
\begin{equation}
 \underline\gamma(w)=
 \frac{\bigl(|w^{\mathrm H}\widehat h|-\epsilon_h\bigr)_+^2}
 {\bigl(|w^{\mathrm H}\widehat g|+\epsilon_g\bigr)^2+\sigma^2}.
 \label{eq:front-robust}
\end{equation}
The triangle inequality gives the numerator lower bound and denominator upper bound. Errors chosen along the complex direction $w$ can reduce the desired amplitude to its permitted minimum and increase the interference amplitude to its permitted maximum. Since the two errors may be chosen independently, both bounds are attained simultaneously. If the desired component can be canceled completely, the worst-case value is zero.

The robust weights can still be found in the two-dimensional response subspace. Write
\begin{equation}
 e_1=\widehat g/\|\widehat g\|,\quad
 c=e_1^{\mathrm H}\widehat h,\quad
 a=\|\widehat h-ce_1\|,\quad
 e_2=(\widehat h-ce_1)/a.
\end{equation}
First suppose $a>0$ and $\widehat g\ne0$. It suffices to search in
$\operatorname{span}\{e_1,e_2\}$. Indeed, suppose the projection of a weight onto this plane has norm $t<1$, and let $A,B$ be the two estimated amplitudes obtained from its normalized projection. Then~\eqref{eq:front-robust} becomes
$(tA-\epsilon_h)_+^2/[(tB+\epsilon_g)^2+\sigma^2]$. Wherever this expression is positive, it increases with $t$: after removing positive factors, the sign of its derivative is that of
$(tB+\epsilon_g)(A\epsilon_g+B\epsilon_h)+A\sigma^2>0$. Removing the component orthogonal to both responses and renormalizing therefore cannot decrease the objective.

Choose the relative phase of the two remaining components so that their desired-signal contributions add coherently. This gives
\begin{align}
 w(\theta)&=e^{\mathrm i\arg c}\cos\theta\,e_1+\sin\theta\,e_2,
 \qquad0\le\theta\le\pi/2,\\
 \underline\gamma(\theta)
 &=\frac{\bigl(|c|\cos\theta+a\sin\theta-\epsilon_h\bigr)_+^2}
 {\bigl(\|\widehat g\|\cos\theta+\epsilon_g\bigr)^2+\sigma^2}.
 \label{eq:front-theta}
\end{align}
The original problem has thus reduced to scalar optimization on a closed interval. The degenerate cases $a=0$ or $\widehat g=0$ are solved directly along the remaining directions. The coding experiments maximize over a uniform grid of 2001 points; the next subsection gives the continuous optimum. If every grid candidate has a zero worst-case guarantee, the nominally optimal weight is included as a fallback candidate.

Under orthogonal training, each response estimation error is
$\mathcal{CN}(0,\sigma^2I_L/r)$, so
$r\|\widehat h-h\|^2/\sigma^2$ has density
$t^{L-1}e^{-t}/(L-1)!$. Denote its $1-\delta/4$ quantile by $q_{L,1-\delta/4}$. The radii
\begin{equation}
 \epsilon_h=\epsilon_g=
 \sqrt{\frac{\sigma^2}{r}q_{L,1-\delta/4}}
 \label{eq:front-radius}
\end{equation}
ensure that all four uncertainty balls at the two receivers hold simultaneously with probability at least $1-\delta$. On this event,~\eqref{eq:front-robust} holds for every candidate weight at once. It therefore also holds for weights selected from the same training data, without any additional independence assumption after selection.

\subsection{Global Optimization of Robust Weights}

The continuous optimization in~\eqref{eq:front-theta} reduces to comparison of the real roots of a quartic polynomial. To distinguish the following real quantities from the complex coefficient $c$, write
$b=|c|$, $d=\|\widehat g\|$, $u=\epsilon_h$, $v=\epsilon_g$, and $\nu=\sigma^2>0$.
Set $t=\tan(\theta/2)\in[0,1]$ and define the real polynomials
\begin{align}
 P(t)&=(b-u)+2at-(b+u)t^2,\\
 Q(t)&=\bigl[(d+v)+(v-d)t^2\bigr]^2+\nu(1+t^2)^2.
\end{align}
The half-angle identities transform the objective into
\begin{equation}
 F(t)=\frac{[P(t)_+]^2}{Q(t)},\qquad Q(t)>0.
 \label{eq:robust-polynomial}
\end{equation}

\medskip\noindent\textbf{Proposition (Finite-candidate characterization of robust observation).}
If the optimum is positive, the global maximum of~\eqref{eq:robust-polynomial} is attained at $t=0$, $t=1$, or a real root satisfying
\begin{equation}
 2P'(t)Q(t)-P(t)Q'(t)=0,\qquad 0<t<1,\quad P(t)>0.
 \label{eq:robust-quartic}
\end{equation}
The polynomial in~\eqref{eq:robust-quartic} has degree at most four. If $P(t)\le0$ throughout the interval, every candidate weight has a zero worst-case guarantee.

\begin{proof}
Continuity of $F$ on the compact interval ensures existence of a maximum. In the interior region where $P>0$, its derivative is
$P(2P'Q-PQ')/Q^2$, so any positive interior maximum must satisfy~\eqref{eq:robust-quartic}.
A boundary point with $P=0$ cannot give a positive maximum. If $p_2$ and $q_4$ are the quadratic coefficient of $P$ and quartic coefficient of $Q$, respectively, then $2P'Q$ and $PQ'$ both have fifth-degree coefficient $4p_2q_4$, which cancels. Comparing the objective at the endpoints and the indicated real roots therefore yields the global maximum.
\end{proof}

This optimum concerns worst-case SINR over independent Euclidean uncertainty balls under a unit-norm linear observation constraint. Neither training duration nor finite-block error probability is part of this objective. The numerical implementation uses floating-point polynomial roots, real-root selection, and objective comparison; it does not use interval arithmetic.

Numerical checks cover 1200 sets of amplitudes, uncertainty radii, and noise parameters, each compared with a 20001-point fine grid and local scalar optimization. A further 200 complex response pairs are each checked against 100 random weights, for 20000 weights in total. The computed candidate solutions are no worse than any of these comparison values, and normalized stationarity residuals are below $1.6\times10^{-12}$. The coding-experiment tables retain the 2001-point grid implementation; the root-finding checks do not remeasure communication performance.

\subsection{Phase Reference and Finite Control Resolution}

Changing the observation weights generally changes the phase of the desired-signal coefficient. If
$a(w_0),a(w_1)\ne0$, phase comparison between the two observation states gives
\begin{equation}
 T_{1\leftarrow0}=\exp\{-\mathrm i[\arg a(w_1)-\arg a(w_0)]\}.
\end{equation}
Multiplication of the new record by this factor brings its desired-signal component to the phase reference of the old record. This phase transport aligns the desired components but leaves the signal, interference, and noise amplitudes of the new record unchanged. The two observations may consequently retain different qualities.

An implementation has access only to $\widehat a(w)=w^{\mathrm H}\widehat h$. If
$|\widehat a(w)|>\epsilon_h$, the true coefficient lies in the complex disk centered at $\widehat a$ with radius $\epsilon_h$. A tangent from the origin to this disk gives the residual phase bound
\begin{equation}
 |\arg a(w)-\arg\widehat a(w)|
 \le\arcsin\frac{\epsilon_h}{|\widehat a(w)|}.
\end{equation}
The phase difference is measured by the shortest distance modulo $2\pi$. This bound no longer applies when the disk contains the origin. The decoder below uses the estimated effective desired-signal coefficient, so residual phase error is included in the measured bit and block error rates.

If the analog front end controls phase but not amplitude independently, admissible weights can be restricted to
$w_a=L^{-1/2}e^{\mathrm i\phi_a}$. With three-bit phase resolution, the four relative phases give $8^3=512$ candidates: a common phase leaves SINR unchanged, so one component phase may be fixed at zero. Exhaustive search gives the global optimum over this finite set. This design uses the same number of analog inputs and sampling chains as continuous amplitude-and-phase control; the admissible observation-state sets differ.

\subsection{Finite-Block Protocol and Performance Measures}

The experiments use $L=4$, $n=256$, and $\sigma^2=1$, with spatial responses
\begin{equation}
 h_i=\sqrt{10}\,v(\xi_i),\quad
 g_i=\sqrt{1000}\,v(\zeta_i),\quad
 v(\xi)_a=\tfrac12 e^{\mathrm i\pi a\xi},\quad a=0,1,2,3.
\end{equation}
For each block, the $\xi_i$ are drawn independently and uniformly from $[-0.45,0.45]$. Set
$\zeta_i=\xi_i\pm d_i$, with equiprobable signs and $d_i$ uniform on $[0.2,0.8]$, and clip the result to
$[-0.95,0.95]$. The resulting spatial formation model has a desired-signal SNR of 10~dB and an interference-to-noise ratio of 30~dB. All schemes share the spatial responses, transmitted messages, and analog noise, allowing paired comparisons block by block.

By default, each training weight is repeated $r=4$ times, for a total of 32 training slots. The remaining 224 slots carry binary modulation symbols. Each user divides 128 information bits into 32 groups and maps each four-bit vector $d=(d_1,d_2,d_3,d_4)$ to
\begin{equation}
 (d_1,d_2,d_3,d_4,
 d_1\oplus d_2\oplus d_4,
 d_1\oplus d_3\oplus d_4,
 d_2\oplus d_3\oplus d_4).
\end{equation}
Transmission uses $0\mapsto+1$ and $1\mapsto-1$. For each seven-symbol record, the receiver searches all 16 desired-user codewords and selects the one with minimum Euclidean distance under the estimated desired response. Interference is treated as noise, and the two receivers decode independently. All schemes use this rule except the local joint-decoding baseline described below.

Eight receiver designs are compared: fixed equal-phase combining; matching only the estimated desired response; the same matched weights with local joint processing of both users' codewords; hard nulling of the estimated interference; substitution of the estimated responses into~\eqref{eq:front-optimum}; maximization of the worst-case guarantee over the uncertainty balls; search over 512 three-bit phase candidates; and an ideal reference with knowledge of the true responses. The ideal reference is charged the same training slots, so its advantage over implementable schemes reflects the effect of response estimation. All other schemes use only local pilots.

The local joint-decoding baseline uses the same pilots and matched weights. For each desired-user codeword $c$, it marginalizes over the 16 equiprobable codewords $d$ of the other user, computing
\begin{equation}
 \ell(c)=\log\sum_{d}\exp\left\{-\frac{\|y-\widehat a c-\widehat b d\|^2}{\sigma^2}\right\},
\end{equation}
and chooses the maximizer of $\ell(c)$. Both $\widehat a$ and $\widehat b$ are estimated from local pilots. Each receiver independently evaluates 256 codeword-pair likelihoods, without receiver cooperation. This baseline improves decoding at fixed observation weights, allowing its benefit to be distinguished from that of changing the weights.

Each experiment contains 4096 independent blocks. Whole-block goodput counts a user's 128-bit payload only when all its bits are recovered correctly. Bit error rate is measured on information bits, and block error rate on complete user messages. Every measure is obtained by comparing transmitted messages with decoder outputs. The protocol has neither an error-detection code nor acknowledgment feedback, so goodput measures correctly recovered payload rather than acknowledged delivery. The true SINR reported in the tables is used only for offline evaluation and is unavailable to both controller and decoder.

\begin{table}[H]
\centering\small\setlength{\tabcolsep}{3.5pt}
\caption{Static-array performance: 128 payload bits per user, 32 pilots, and 256 channel uses per frame; 4096 frames per policy. SINR statistics pool the two receivers.}
\label{tab:front-main}
\begin{tabular}{lrrrrr}
\toprule
Policy & \shortstack{Median\\SINR} & 5th pct. & BER & \shortstack{Joint\\BLER} & Goodput\\
 & dB & dB & \% & \% & bit/use\\
\midrule
Fixed weights & -11.25 & -30.57 & 38.8603 & 97.803 & 0.1484\\
Matched filter & -8.97 & -17.06 & 32.6625 & 93.750 & 0.2570\\
Matched + joint decoding & -8.97 & -17.06 & 0.2335 & 3.931 & 0.9799\\
Estimated null & 8.55 & 5.68 & 0.0009 & 0.122 & 0.9994\\
SINR-optimal weights & 8.56 & 5.68 & 0.0009 & 0.122 & 0.9994\\
Robust weights & 8.55 & 5.66 & 0.0997 & 0.562 & 0.9972\\
3-bit phase control & 6.24 & 0.04 & 0.4143 & 9.839 & 0.9497\\
Known response & 9.71 & 7.11 & 0.0000 & 0.000 & 1.0000\\
\bottomrule
\end{tabular}
\end{table}
\begin{table}[H]
\centering\small\setlength{\tabcolsep}{3.5pt}
\caption{Pilot selection and independent validation under a 1\% joint block-error constraint. The three candidates were fixed after separate screening; the bounds hold simultaneously with confidence at least 95\%.}
\label{tab:front-training}
\begin{tabular}{lrrrrr}
\toprule
Policy & Pilots & \shortstack{Payload\\per user} & \shortstack{Joint\\BLER} & \shortstack{95\%\\bound} & Goodput\\
 & & bit & \% & \% & bit/use\\
\midrule
Estimated null & 32 & 128 & 0.073 & 0.228 & 0.9996\\
SINR-optimal weights & 32 & 128 & 0.122 & 0.300 & 0.9994\\
Robust weights & 32 & 128 & 0.562 & 0.867 & 0.9972\\
\bottomrule
\end{tabular}
\end{table}

\begin{figure}[htbp]
\centering\includegraphics[width=.95\textwidth]{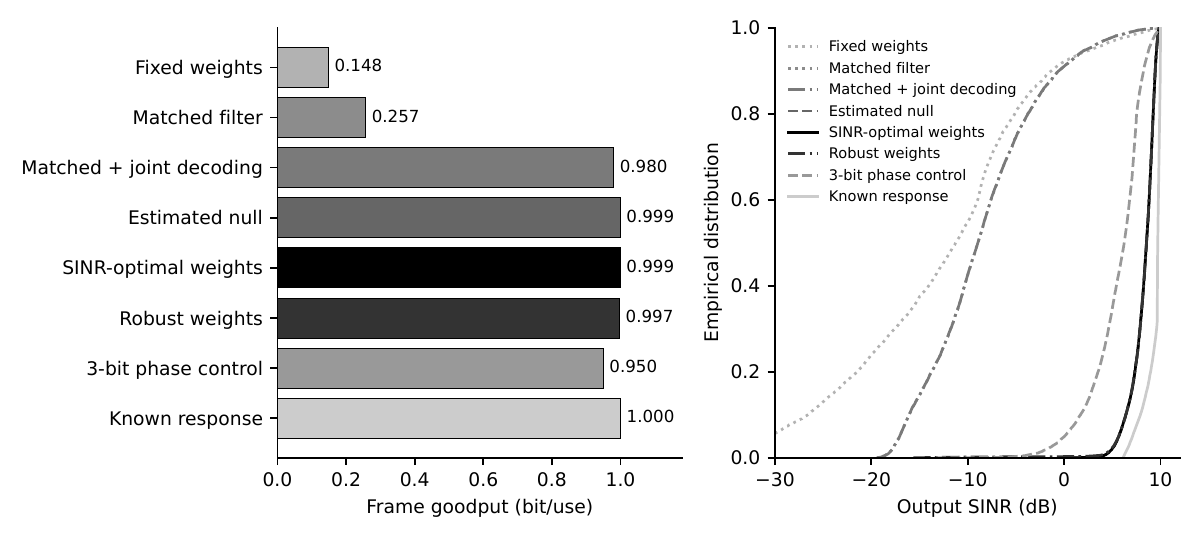}
\caption{Performance of observation designs with a single sampling chain. All schemes use the same training length and code. Left: whole-block goodput. Right: empirical distribution of output SINR. The ideal reference uses the true responses.}
\end{figure}

\subsection{Training Overhead and Reliability}

With the default 32 training slots, optimization using estimated responses raises whole-block goodput from 0.2570 for desired-response matching to 0.9994 bits per slot, a factor of approximately 3.89. Median output SINR rises from $-8.97$~dB to $8.56$~dB. With the sampling chain count and code unchanged, the spatial weights alter how the record is formed: average interference power falls from 130.51 to 0.2454, while average desired-signal power changes from 9.30 to 8.33.

The local joint-decoding baseline achieves 0.9799 correctly recovered payload bits per slot, with a joint block error rate of 3.93\%. Observation-weight optimization increases goodput by approximately 1.99\% to 0.9994 bits per slot and reduces joint block error rate to 0.12\%. It also compares only 16 desired-user candidates per seven-symbol codeword, whereas the joint baseline processes 256 codeword pairs. Candidate count measures the search size; runtime depends on the implementation. The same matched weights give the same SINR under both decoders but different error rates. Thus SINR alone determines neither finite-code performance nor the capacity of the complete protocol.

Hard nulling based on the estimated responses gives the same whole-block recovery results. Its median SINR differs from full SINR optimization by only about 0.005~dB. In this strong-interference regime, spatial separation accounts for most of the gain, with little additional benefit from SINR optimization over hard nulling. The three-bit phase design achieves whole-block goodput of 0.9497. Phase-only control therefore captures much of the improvement, although lower-tail SINR and block error rate remain limiting factors.

The robust design maximizes worst-case SINR within the uncertainty balls. Its joint block error rate is 0.56\%, compared with 0.12\% for optimization using estimated responses. Worst-case SINR and average block error are different criteria; improvement in the former need not improve the latter. The four confidence balls at the two receivers simultaneously cover the true responses in approximately 95.14\% of samples, consistent with the 95\% guarantee. Samples outside these balls are included in all error statistics.

Training length is also selected over $r\in\{1,2,4,8\}$, with total blocklength fixed at 256. Each scheme is first screened on 1024 independent blocks, after which its selected configuration is validated on 4096 further blocks. When expected goodput is the sole objective, optimization using estimated responses selects $r=2$, reducing pilot use to 16 slots and increasing payload to 136 bits per user. Independent validation gives goodput of 1.0551 bits per slot, but joint block error rate rises to 1.37\%. Fewer pilots provide more data slots while increasing the decoding risk due to estimation error.

For a target joint block error rate of at most 1\%, configurations with screening error rates no greater than 0.5\% are retained, and the one with the highest goodput is then fixed. Optimization using estimated responses, hard nulling, and robust design all select $r=4$. The three fixed candidates are evaluated using simultaneous one-sided binomial bounds at 95\% confidence; the bounds appear in Table~\ref{tab:front-training}. Configuration selection thus includes the observation weights, training directions, and training duration, with duration optimized only over the four specified repetition counts.

\begin{figure}[htbp]
\centering\includegraphics[width=.95\textwidth]{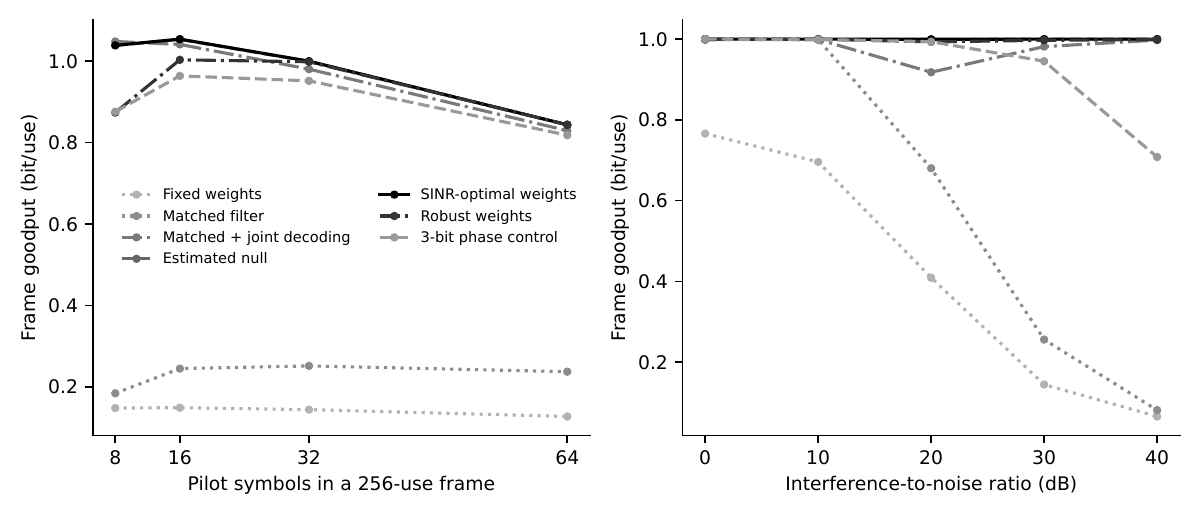}
\caption{Training resources and interference strength. Left: independent screening samples show the estimation error caused by insufficient training and the loss of data slots caused by excessive training. Right: interference power varies with training fixed at 32 slots. All curves use actual encoding and local decoding.}
\end{figure}

These results assume a linear front end, responses that remain fixed within a block, and the stated noise law. Spatial separability determines which directions are available; training error determines how accurately they can be identified. Together they limit the receiver channels that can be formed. Saturation before combining invalidates the linear model, and unmodeled calibration bias may invalidate the guarantees. RF insertion loss, quantization, and switching transients require additional component models. Time-varying responses further require causal tracking.
\section{Causal Observation Design for Time-Varying Channels}
\label{sec:dynamic-observation}

Motion of the spatial responses and drift in array amplitudes and phases change the receiver channel. Fixed weights then gradually lose the separation achieved during training. A time-varying design must predict the effective responses from past pilots and update its weights within the budgets for training time, scalar observations, and switching. Prediction error contains both estimation noise and response curvature, each of which depends on the extrapolation horizon.

\subsection{Effective Responses and Array Uncertainty}

For each receiver, let
\begin{equation}
 r(t)=D(t)\bigl[h(t)u_i(t)+g(t)u_j(t)+\nu(t)\bigr],
 \qquad y(t)=w(t)^{\mathrm H}r(t),\quad \|w(t)\|=1,
 \label{eq:dynamic-field}
\end{equation}
where $D(t)$ is an unknown diagonal matrix of complex gains satisfying
$0<a_{\min}\le |D_{aa}(t)|\le a_{\max}$, and the noise samples
$\nu(t)\sim\mathcal{CN}(0,\sigma^2I_L)$ are independent across time. Define the effective responses
\begin{equation}
 F(t)=D(t)h(t),\qquad G(t)=D(t)g(t),\qquad
 v_0=\sigma^2a_{\max}^2.
 \label{eq:effective-dynamic}
\end{equation}
The conditional law of the record is determined by $w^{\mathrm H}F$, $w^{\mathrm H}G$, and the noise variance, which satisfies
$\sigma^2\|D^{\mathrm H}w\|^2\le v_0$. The receiver estimates $F$ and $G$ through a single scalar observation chain, without prior knowledge of $D$, $h$, or $g$.

When the unknown $D$ is constant, pilots acquired through the same front end already incorporate its effect, allowing direct estimation of the effective responses. The decomposition into array gain and propagation response is generally nonunique. For any diagonal unitary matrix $U$, the transformation
$(D,h,g)\mapsto(DU^{\mathrm H},Uh,Ug)$ preserves $F$, $G$, and $DD^{\mathrm H}$ and hence the probability law of every scalar record. The effective responses are therefore sufficient for observation design.

Since $D$ is invertible, $F\notin\operatorname{span}(G)$ is equivalent to
$h\notin\operatorname{span}(g)$. More precisely,
\begin{equation}
 \operatorname{dist}\bigl(F,\operatorname{span}(G)\bigr)
 \ge a_{\min}\operatorname{dist}\bigl(h,\operatorname{span}(g)\bigr).
 \label{eq:dynamic-separation}
\end{equation}
Apply $\|D(h-cg)\|\ge a_{\min}\|h-cg\|$ for each complex $c$ and then take the infimum. Thus an invertible fixed amplitude-and-phase distortion preserves spatial separability, although it changes the separating directions and the retained signal strength. Training must identify the distorted effective responses.

\subsection{Causal Response Prediction}

As before, one user transmits known pilots while the other transmits zero. For any of the four effective responses
$f\in\{F_1,G_1,F_2,G_2\}$, the observations are
\begin{equation}
 z_j=q_j^{\mathrm H}f(t_j)+\eta_j,\qquad
 \|q_j\|=1,\quad t_j\le\tau.
 \label{eq:dynamic-pilot}
\end{equation}
Pilot directions and sampling times are prescribed in advance, and $\tau$ is the end of the current training round. The physical response may be random, but its exogenous trajectory is assumed independent of receiver thermal noise. Conditional on that trajectory, the $\eta_j$ are independent circularly symmetric complex Gaussian variables with variances at most $v_0$. The probability calculations below condition on the trajectory; the receiver itself still has access only to past pilots.

Write $s_j=t_j-\tau$, and form $\Phi$ with row $j$ equal to $[q_j^{\mathrm H},s_jq_j^{\mathrm H}]$. Assume $\rank\Phi=2L$. The local affine estimate and its extrapolation are
\begin{equation}
 \widehat\vartheta=(\Phi^{\mathrm H}\Phi)^{-1}\Phi^{\mathrm H}z,
 \qquad
 \widehat f(t)=B_t\widehat\vartheta,\qquad
 B_t=[I_L,(t-\tau)I_L],\quad t\ge\tau.
 \label{eq:dynamic-predictor}
\end{equation}
The first parameter block estimates the current effective response; the second estimates its rate of change. A full-column-rank $\Phi$ is obtained by taking $L$ orthogonal directions and observing each at two or more distinct times. All observations are acquired sequentially through the same sampling chain, and training for both users' responses is charged to the total duration.

Assume the effective response obeys the curvature bound
\begin{equation}
 \|f''(v)\|\le A_f
 \label{eq:dynamic-curvature}
\end{equation}
throughout the training interval and the subsequent prediction interval. The regularity bound $A_f$ is specified before design. Put $P=(\Phi^{\mathrm H}\Phi)^{-1}\Phi^{\mathrm H}$ and define
\begin{align}
 V_t&=B_t(\Phi^{\mathrm H}\Phi)^{-1}B_t^{\mathrm H},\\
 \beta_f(t)&=\frac{A_f}{2}
 \left[(t-\tau)^2+\|B_tP\|_2
 \left(\sum_j s_j^4\right)^{1/2}\right].
 \label{eq:dynamic-bias}
\end{align}
The quantity $\beta_f$ bounds the deterministic remainder of the affine approximation. Closely spaced pilots amplify noise in the slope estimate, whereas a long prediction horizon increases the curvature remainder. Both errors must be controlled in choosing the predictor.

\medskip\noindent\textbf{Proposition (Finite-horizon prediction guarantee).}
Suppose there are $K$ prescribed data-readout times. Let $q_{L,1-\alpha}$ denote the $1-\alpha$ quantile of the distribution with density
$x^{L-1}e^{-x}/(L-1)!$, and set
\begin{equation}
 \epsilon_f(t)=\beta_f(t)+
 \sqrt{v_0\lambda_{\max}(V_t)q_{L,1-\delta/(4K)}}.
 \label{eq:dynamic-radius}
\end{equation}
Then, with probability at least $1-\delta$, all four effective responses at the two receivers satisfy
$\|f(t)-\widehat f(t)\|\le\epsilon_f(t)$ simultaneously at these $K$ times.

\begin{proof}
Let $\vartheta=[f(\tau)^{\mathsf T},f'(\tau)^{\mathsf T}]^{\mathsf T}$.
Taylor's formula gives $z=\Phi\vartheta+b+\eta$, with
$|b_j|\le A_fs_j^2/2$, and also
$f(t)=B_t\vartheta+e_t$, with $\|e_t\|\le A_f(t-\tau)^2/2$.
Hence
\[
 \widehat f(t)-f(t)=B_tPb-e_t+B_tP\eta.
\]
The first two terms have combined norm bounded by~\eqref{eq:dynamic-bias}. The last term is a zero-mean complex Gaussian vector whose covariance satisfies
\[
 C_t=B_tP\operatorname{Cov}(\eta)P^{\mathrm H}B_t^{\mathrm H}
 \preceq v_0V_t.
\]
It can be written as $C_t^{1/2}\zeta$, with $\zeta\sim\mathcal{CN}(0,I_L)$, so that
$\|C_t^{1/2}\zeta\|^2\le v_0\lambda_{\max}(V_t)\|\zeta\|^2$.
The squared norm $\|\zeta\|^2$ has the density stated in the proposition. Assigning tail probability $\delta/(4K)$ to each response at each time and applying the union bound proves the result. Prediction errors at different times may be correlated; their independence is not required.
\end{proof}

The proposition assumes training directions and times fixed in advance. If regression samples are chosen adaptively from observed noise, the resulting dependence between selection and noise requires a separate argument. Subsequent data weights may depend on these pilots because the uncertainty balls cover the entire response vectors. To guarantee every data readout between updates, all such times must be included in $K$.

\subsection{Robust Weights from Predicted Responses}

On the joint event above, every unit-norm weight satisfies
\begin{equation}
 \gamma_t(w)\ge\underline\gamma_t(w)
 :=\frac{\bigl(|w^{\mathrm H}\widehat F(t)|-\epsilon_F(t)\bigr)_+^2}
 {\bigl(|w^{\mathrm H}\widehat G(t)|+\epsilon_G(t)\bigr)^2+v_0}.
 \label{eq:dynamic-robust}
\end{equation}
Maximizing the right-hand side at the permitted update times gives causal weights depending only on past pilots. Equation~\eqref{eq:dynamic-robust} is a lower bound on SINR in the physical model. A common $D$ couples the desired response, interference response, and noise, whereas independent uncertainty balls enlarge the uncertainty set. Their extreme combinations need not arise from a single physical state, so the guarantee may be conservative.

The optimization again reduces to one dimension. From the current predicted responses, form
$e_1=\widehat G/\|\widehat G\|$, $c=e_1^{\mathrm H}\widehat F$,
$a=\|\widehat F-ce_1\|$, and $e_2=(\widehat F-ce_1)/a$. In the nondegenerate case, it remains to solve
\begin{equation}
 \max_{0\le\theta\le\pi/2}
 \frac{\bigl(|c|\cos\theta+a\sin\theta-\epsilon_F(t)\bigr)_+^2}
 {\bigl(\|\widehat G\|\cos\theta+\epsilon_G(t)\bigr)^2+v_0}.
 \label{eq:dynamic-one-dimensional}
\end{equation}
The corresponding weight is $e^{\mathrm i\arg c}\cos\theta\,e_1+\sin\theta\,e_2$.
Projection and phase alignment follow the static argument. A zero optimum means that this uncertainty set yields no positive uniform guarantee; it does not imply zero SINR in the true channel.

With a finite switching budget, the current weights may be retained until their guarantee falls below a specified target and new weights can improve it. The training schedule remains fixed, and each switch is charged to the protocol's resource budget. This threshold policy respects the switching constraint; its finite-block performance must still be evaluated together with encoding and decoding.

\subsection{Phase Error and Prediction Horizon}

Let $a_t=w_t^{\mathrm H}F(t)$ and $\widehat a_t=w_t^{\mathrm H}\widehat F(t)$. If
$|\widehat a_t|>\epsilon_F(t)$, the complex disk containing the true coefficient excludes the origin, and hence
\begin{equation}
 d_{\mathbb S^1}\bigl(\arg a_t,\arg\widehat a_t\bigr)
 \le\arcsin\frac{\epsilon_F(t)}{|\widehat a_t|},
 \label{eq:dynamic-phase}
\end{equation}
where $d_{\mathbb S^1}$ is the shortest angular distance modulo $2\pi$. A phase reference between two times constructed from the predicted coefficients has residual error at most the sum of the two endpoint bounds. This error includes both the temporal evolution of the physical response and changes in the observation weights. Comparing two observation states at the same time changes only the weights. The phase guarantee fails when the disk contains the origin.

A curvature bound follows directly from bounds on array amplitude, phase, and propagation response. Since
\begin{equation}
 F''=D''h+2D'h'+Dh'',
\end{equation}
the assumptions $\|h\|\le H$, $\|h'\|\le v_h$, $\|h''\|\le a_h$,
$\|D'\|\le d_1$, and $\|D''\|\le d_2$ permit the choice
\begin{equation}
 A_F=d_2H+2d_1v_h+a_{\max}a_h.
 \label{eq:dynamic-curvature-budget}
\end{equation}
The interference response is treated in the same way. Writing each gain as $d_a=\rho_a e^{\mathrm i\phi_a}$ gives
\begin{align}
 |d_a'|&\le |\rho_a'|+\rho_a|\phi_a'|,\\
 |d_a''|&\le |\rho_a''|+2|\rho_a'||\phi_a'|
 +\rho_a\bigl(|\phi_a''|+|\phi_a'|^2\bigr),
\end{align}
which converts component drift limits into bounds needed for prediction. Constant unknown array errors correspond to $d_1=d_2=0$. Faster variation shortens the horizon over which reliable extrapolation is possible.

The curvature assumption is indispensable. Two trajectories may agree at every past pilot time yet have arbitrarily different phases at the next data time. No controller receiving the same history can distinguish them. Causal prediction guarantees therefore depend on temporal regularity and on training records whose acquisition is charged to the resource budget. Front-end saturation, abrupt response changes, and motion beyond the assumed curvature range lie outside the theorem's conditions.
\section{Finite-Blocklength Performance of Time-Varying Observation Channels}

\subsection{Channel Model and Causal Reception Protocol}

Consider a receiver with four analog inputs and one sampling chain. The block length is 1024, the normalized desired and interfering powers are 10 and 1000, and thermal noise variance is 1 in each input. Initial propagation directions have the same distribution as in the static experiment. With normalized time $s=t/1023$, each response direction evolves as $\xi(s)=\xi_0+vs$, where $v$ is independently uniform on $[-0.2,0.2]$. Initial array-gain magnitudes are uniform on $[0.9,1.1]$ and remain constant throughout the block. Input phases follow $\phi_a(s)=\phi_{a,0}+\omega_as$, with initial phases uniform on $[-0.25,0.25]$ and $\omega_a$ uniform on $[-0.4,0.4]$, all in radians. The simulation generates the effective responses and noise slot by slot, whereas the controller uses only the stated parameter ranges and past pilots. Time is normalized by block length; the results are not yet tied to a particular carrier frequency, speed, or duration in seconds.

The first 128 slots provide initial training. Each subsequent period begins with 16 training slots followed by data. Training again uses the four Fourier directions in sequence: one user sends four pilots while the other is silent, then their roles are exchanged. Two such rounds occupy 16 slots. The period is selected from $\{64,128,256\}$. Data use the preceding section's seven-symbol code carrying four information bits. Slots insufficient for a complete codeword at the end of a period carry zero symbols from both users and count toward the total duration. Every scheme follows a schedule agreed before transmission; pilot scheduling requires no receiver feedback.

The affine predictor fits timestamped scalar records directly, using pilots acquired during the preceding 256 slots. Relative times are divided by 128 solely to improve matrix conditioning, without changing the predictor. Before each seven-symbol codeword, past pilots are used to predict both effective responses at every symbol time. The weight is designed from the midpoint prediction and held fixed for seven slots. The decoder uses the individually predicted complex coefficients at each symbol time, including their changing phase reference. All implementable schemes use the known noise upper bound $v_0=1.21$; the actual noise variance $\|D^{\mathrm H}w\|^2$ is available only for offline evaluation.

Six observation schemes are compared. The frozen-weight scheme designs its data weight once after initial training but continues to update its decoding coefficients, separating weight adaptation from coefficient updating. Periodic re-estimation fits constant responses to the latest 16 pilots and holds both the data weight and estimated coefficients throughout the period. Predictive SINR maximization redesigns the weight for each codeword from the predicted responses; predictive hard nulling cancels the predicted interference response. Predictive matched combining uses the same affine predictor as predictive SINR maximization but aligns the weight with the desired response, so these two schemes differ only in their weight-design criterion. The true-response reference uses the current effective responses and actual diagonal noise covariance, designing its weight from the true midpoint responses. It retains the same pilot and codeword schedule; the performance difference measures the value of response information.

All schemes use the same local marginal-likelihood decoder. Each desired-user codeword is evaluated with all 16 interfering codewords, for a total of 256 pairs. With time-varying coefficients, the score is
\begin{equation}
 \ell(c)=\log\sum_d\exp\left[-\frac1{v_0}
 \sum_{q=1}^7|y_q-\widehat a_qc_q-\widehat b_qd_q|^2\right].
 \label{eq:dynamic-decoder}
\end{equation}
The cross term in the squared expansion is
$2\sum_q\operatorname{Re}(\widehat a_q^*\widehat b_q)c_qd_q$ and must be evaluated with the coefficients at each symbol time. The decoder substitutes estimated responses for the true responses rather than integrating over their full posterior. The true-response reference uses the same noise-upper-bound parameter, so its performance also depends on this prescribed decoder. Each receiver recovers its own message without exchanging records.

These weights maximize nominal SINR at the midpoint of each codeword. The posterior recursion of Section~\ref{sec:policy-optimum} instead minimizes terminal message error; the objectives and domains of optimality differ. The curvature-based confidence balls of Section~\ref{sec:dynamic-observation} could be used to construct robust policies. This experiment uses nominal predicted weights and measures block error from actual decoding outcomes.

\subsection{Reliability, Goodput, and Control Overhead}

Each of the two main schemes selects its training period using 384 independent blocks; both choose 64. With this period fixed, all six schemes are compared on a further 2048 blocks, sharing physical trajectories, messages, and noise for paired comparisons. Each block contains 352 pilots, 588 coded-data slots, and 84 idle slots, carrying 336 information bits per user. Training occupies 34.375\% of the block and idle time 8.203\%; the nominal sum rate is only 0.65625 bit/slot. Goodput uses the full block length, including all overhead. Period selection is limited to the three candidates and is conditional on the chosen predictor.

Writing $k$ for each user's payload, whole-frame sum goodput is
\begin{equation}
 G_\Sigma=\frac{k}{n}\sum_{i=1}^2\Pr\{\widehat M_i=M_i\}.
\end{equation}
This metric credits each user's entire payload when it is recovered correctly. Failure by one user does not remove credit for the other's successfully recovered payload. Joint error, in contrast, occurs when at least one user's whole-frame message is decoded incorrectly.

\begin{table}[H]
\centering\small\setlength{\tabcolsep}{3.5pt}
\caption{Dynamic-array performance over 2048 frames per policy. Goodput is measured in bit/use; BER and joint frame-error rates are percentages. All policies use the same local joint decoder over 256 codeword pairs.}
\label{tab:dynamic-main}
\begin{tabular}{lrrrrr}
\toprule
Policy & \shortstack{Frame\\goodput} & \shortstack{Codeword\\goodput} & BER & \shortstack{Joint\\error} & SINR (dB)\\
\midrule
Frozen weights & 0.5053 & 0.6440 & 0.951 & 40.33 & 0.71\\
Periodic re-estimation & 0.5898 & 0.6443 & 0.921 & 19.38 & 7.40\\
Predictive SINR optimum & 0.5883 & 0.6449 & 0.876 & 19.87 & 7.67\\
Predictive null & 0.5885 & 0.6449 & 0.873 & 19.82 & 7.67\\
Predictive matched filter & 0.5342 & 0.6526 & 0.284 & 33.74 & -10.16\\
Known response & 0.6279 & 0.6519 & 0.312 & 8.45 & 9.72\\
\bottomrule
\end{tabular}
\end{table}
\begin{table}[H]
\centering\small\setlength{\tabcolsep}{3.5pt}
\caption{Output powers, phase error, and weight changes. Powers use the normalization in the text. Phase error is the 95th percentile of the desired-channel coefficient estimation error, in degrees. The switch bound is per receiver and includes pilot-direction changes.}
\label{tab:dynamic-resources}
\begin{tabular}{lrrrrr}
\toprule
Policy & Signal & Interference & Noise & Phase error & Switch bound\\
\midrule
Frozen weights & 7.870 & 20.854 & 1.005 & 21.59 & 365\\
Periodic re-estimation & 7.627 & 0.580 & 1.004 & 21.73 & 365\\
Predictive SINR optimum & 7.672 & 0.477 & 1.004 & 21.28 & 435\\
Predictive null & 7.669 & 0.476 & 1.004 & 21.32 & 435\\
Predictive matched filter & 8.805 & 169.586 & 1.011 & 17.67 & 435\\
Known response & 8.441 & 0.052 & 0.993 & 0.00 & 435\\
\bottomrule
\end{tabular}
\end{table}
\begin{table}[H]
\centering\small\setlength{\tabcolsep}{3.5pt}
\caption{Training-period selection from 384 frames per candidate. Maximizing frame goodput separately for each policy selects a period of 64 channel uses in both cases. Screening and validation samples are independent.}
\label{tab:dynamic-period}
\begin{tabular}{rrrrrr}
\toprule
Period & Pilots & Idle uses & \shortstack{Payload\\per user} & \shortstack{Predictive\\optimum} & \shortstack{Predictive\\matched}\\
\midrule
64 & 352 & 84 & 336 & 0.5887 & 0.5477\\
128 & 240 & 0 & 448 & 0.4193 & 0.3760\\
256 & 192 & 6 & 472 & 0.0000 & 0.0000\\
\bottomrule
\end{tabular}
\end{table}

Predictive SINR maximization achieves whole-frame sum goodput of 0.58832 bit/slot, compared with 0.53416 for predictive matched combining, an increase of approximately 10.14\%. The two schemes share the response predictor, training schedule, and decoder, so the paired difference isolates the choice of observation direction. The mean paired blockwise difference is 0.05415 bit/slot, with descriptive 95\% normal-approximation interval $[0.04353,0.06478]$. Relative to frozen initial weights, goodput increases by approximately 16.42\%, reflecting the benefit of adapting the observation direction to changing responses.

Periodic re-estimation achieves 0.58976 bit/slot. For predictive SINR maximization minus periodic re-estimation, the paired difference interval is $[-0.00481,0.00193]$. These data do not resolve a performance difference. Predictive hard nulling also gives nearly the same goodput. Periodic re-estimation and affine prediction differ in both estimation method and weight-update frequency, so their comparison reflects both changes.

The mean information-bit error rate under predictive SINR maximization is 0.876\%, higher than the matched scheme's 0.284\%. Its goodput counted by four-bit codewords is also lower, despite its higher whole-frame goodput. These orderings depend on how errors are distributed across frames: many errors in a few frames and a few errors in many frames affect bit error and whole-frame recovery differently. The two metrics must therefore be considered separately; SINR alone does not determine their ranking. In the main experiment, predictive SINR maximization still has a joint whole-frame error rate of 19.87\%, with a one-sided 95\% binomial upper bound of approximately 21.38\%, well above the 1\% reliability target.

The two update mechanisms also incur different control costs. Periodic re-estimation requires only 14 data-weight designs, whereas codeword-by-codeword prediction requires 84. Their switching upper bounds, including training directions, are 365 and 435, respectively. Even frozen data weights must be restored after each periodic pilot sequence; these returns are included in the switching count. Every scheme acquires 1024 scalars per receiver. Whole-frame delivery per observation is one-half of the tabulated sum goodput. Converting read and switch counts into joule efficiency requires an RF hardware energy model. Goodput is scored against the transmitted messages and includes neither error detection nor acknowledgment and retransmission delay.

\begin{figure}[H]
 \centering\includegraphics[width=.97\textwidth]{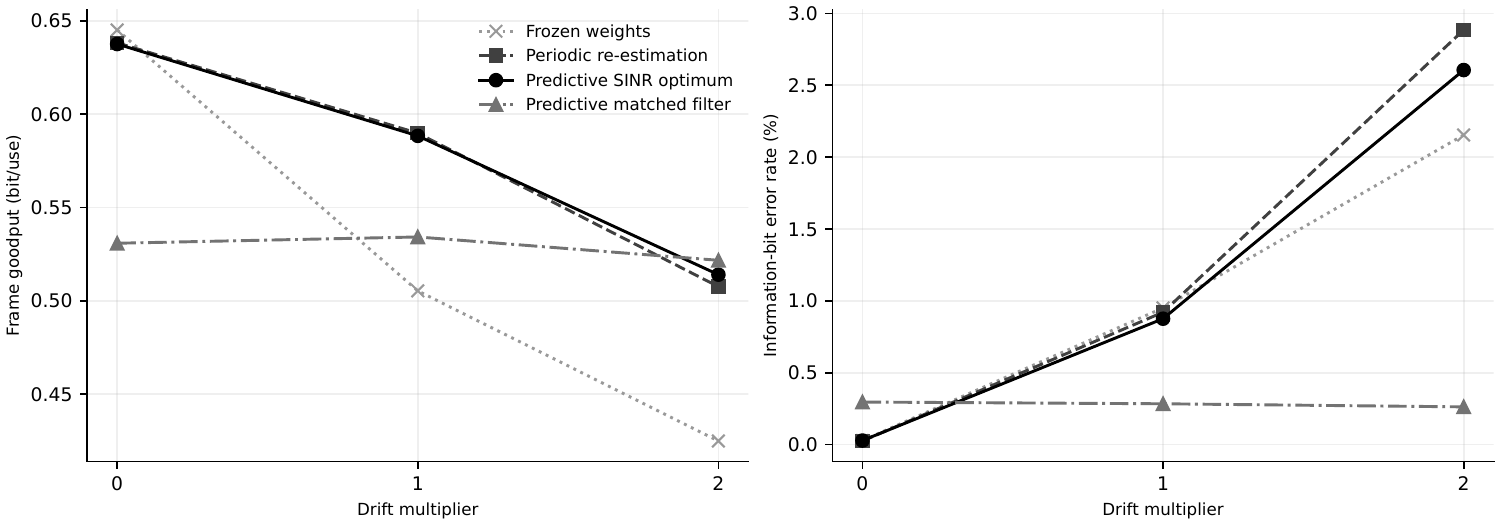}
 \caption{Performance with training period 64 and both propagation-direction and array-phase drift ranges scaled by 0, 1, and 2. The first two cases use 2048 blocks each; the last uses 512. The period is not reselected. The two metrics give different rankings for the same fixed finite code.}
 \label{fig:dynamic-tradeoff}
\end{figure}

Without drift, frozen initial weights suffice, and repeated prediction introduces additional estimation fluctuations. At twice the nominal drift, predictive SINR maximization yields whole-frame goodput of 0.51398, compared with 0.52167 for matched combining, so its point-estimate advantage disappears. This smaller sample does not establish a significant difference. Drift also changes the angle between the desired and interfering responses and may degrade spatial separation in its lower tail; performance loss therefore reflects both tracking error and changing physical separability. The finite-policy posterior recursion and the algebraic weight optimization for given responses are each optimal for their stated objectives. The results in this section compare specific reception protocols on a time-varying unknown channel; the globally optimal policy for complete message recovery in that model remains undetermined.

\section{Observation-Policy Capacity of a Binary Interference Channel}

Consider a binary interference channel with independent messages, separate encoders, and separate decoders. There is no message sharing, transmitter feedback, or receiver cooperation. The physical ports remain unchanged as the admissible observation policies expand from fixed-port operation to open-loop switching and then causal selection. For this finite-field model, all three capacity regions can be determined exactly.

All operations are over $\mathbb F_2$. The independent uniform messages are
$M_i\in\{1,\ldots,N_i\}$; write $k_i=\log_2N_i$ and let
$U_i^n=f_i(M_i)$ be the transmitted sequences. The equiprobable state
$H\in\{0,1\}$ is independent of the messages and random seeds and remains constant throughout a block of length $n$. It represents the block environment and is initially unknown at every terminal.

For $j\ne i$, receiver $i$ has physical input vector and ports
\begin{align}
 V_{i,t}&=(U_{i,t},U_{j,t})^{\mathsf T},\\
 Y_{i,t}^{(s)}&=(1,H\oplus s)V_{i,t}
 =U_{i,t}\oplus(H\oplus s)U_{j,t},\qquad s\in\{0,1\}.
 \label{eq:poda-port}
\end{align}
At each time, the receiver reads one bit from its selected port $S_{i,t}$. If $S_{i,t}=H$, the output is its own user's symbol; otherwise, it is the sum modulo two of the two users' symbols. The unselected port supplies no record.

A fixed-port policy holds $S_{i,t}=s_i$ throughout the block. An open-loop policy permits any predetermined sequence $S_i^n$, independent of the observations. A causal policy may choose $S_{i,t}$ from its own past actions and outputs. In all three classes, the final decoder uses the complete local record and the known pilots. Although an open-loop action sequence is fixed in advance, its decoder may use state information obtained from the pilots.

The joint block error probability is
\begin{equation}
 \epsilon_n=\Pr\{(\widehat M_1,\widehat M_2)\ne(M_1,M_2)\}.
 \label{eq:poda-error}
\end{equation}
The probability averages over messages, the state, and admissible random seeds. For fixed $0\le\epsilon<1/2$, the $\epsilon$-capacity criterion is
$\limsup_n\epsilon_n\le\epsilon$. Under the maximal joint error criterion, the error probability is first computed for each message pair and then maximized over message pairs; it still averages over the equiprobable state $H$ and independent seeds. All regions below use joint block error.

\medskip\noindent
\textbf{Theorem 1 (Capacity regions of the binary port channel).}
For every fixed $\epsilon<1/2$, the capacity regions of the three policy classes are
\begin{align}
 \mathcal C_{\rm fix}
 &=\{(R_1,R_2)\ge0:R_1+R_2\le1\},\label{eq:poda-fixed-region}\\
 \mathcal C_{\rm ol}
 &=\{(R_1,R_2)\ge0:2R_1+R_2\le2,\ R_1+2R_2\le2\},
 \label{eq:poda-open-region}\\
 \mathcal C_{\rm act}&=[0,1]^2.\label{eq:poda-active-region}
\end{align}
These regions are the same under average and maximal joint error. They also remain unchanged under the hard constraint of at most one port switch per receiver per block. The corresponding sum capacities are $1$, $4/3$, and $2$.

For $1/2\le\epsilon<1$, all three $\epsilon$-capacity regions equal $[0,1]^2$. Thus the equiprobable two-state mixture has a discontinuity in its capacity region at the error threshold $1/2$.

\subsection{A Finite-Blocklength Outer Bound for Open-Loop Observation}

\noindent\textbf{Lemma 1 (Two-state intersection bound).}
For any length-$n$ open-loop code with $\epsilon_n<1/2$, let
$\delta_n=1-2\epsilon_n>0$. Then
\begin{align}
 N_1^2N_2&\le\frac{2^{2n}}{\delta_n^2}, &
 N_1N_2^2&\le\frac{2^{2n}}{\delta_n^2}.\label{eq:poda-count-bound}
\end{align}
The encoders need not be injective.

\begin{proof}
First consider a deterministic code. Let $G_h$ be the set of message pairs decoded correctly by both receivers in state $h$. The codebooks and predetermined action sequences are the same in both states. Since the state is equiprobable and the messages are uniform,
\[
 |G_0|+|G_1|=2(1-\epsilon_n)N_1N_2.
\]
Hence $G=G_0\cap G_1$ satisfies $|G|\ge\delta_nN_1N_2$.

For a fixed message $m_1$, put
\[
 g_{m_1}=|\{m_2:(m_1,m_2)\in G\}|.
\]
Let $a_{m_1}$ and $b_{m_1}$ be the numbers of distinct receiver-1 outputs produced by this row of message pairs in states 0 and 1, respectively. The open-loop actions are identical in the two states, so at each coordinate the two outputs are $U_1$ and $U_1\oplus U_2$, in some order. Given $m_1$, the pair of output sequences therefore determines $U_2^n$.

If two distinct values of $m_2$ in this row produced the same $U_2^n$, receiver 2 would also have the same output for both in either fixed state, contradicting correct decoding. The map from $m_2$ in this row of $G$ to the ordered pair of state-dependent outputs is consequently injective, and
\[
 g_{m_1}\le a_{m_1}b_{m_1}.
\]
Receiver 1 decodes correctly on $G$, so within each state the output sets for distinct $m_1$ are disjoint. Thus
\[
 \sum_{m_1}a_{m_1}\le2^n,\qquad
 \sum_{m_1}b_{m_1}\le2^n.
\]
Using $g_{m_1}\le N_2$ and the Cauchy--Schwarz inequality gives
\begin{align*}
 \delta_nN_1\sqrt{N_2}
 &\le\frac{|G|}{\sqrt{N_2}}
 =\sum_{m_1}\frac{g_{m_1}}{\sqrt{N_2}}\\
 &\le\sum_{m_1}\sqrt{g_{m_1}}
 \le\sum_{m_1}\sqrt{a_{m_1}b_{m_1}}\\
 &\le\sqrt{\Bigl(\sum_{m_1}a_{m_1}\Bigr)
                  \Bigl(\sum_{m_1}b_{m_1}\Bigr)}
 \le2^n.
\end{align*}
Squaring proves the first inequality; interchanging the users proves the second.

For a randomized code, all common and private seeds are assumed independent of the messages and $H$. Fix a realization of the seeds whose average joint error is no greater than the randomized average. The resulting deterministic code has the same message sizes, so the counting bound applies. Since maximal joint error is at least average joint error, the same outer bound holds under the maximal error criterion.
\end{proof}

Taking logarithms yields the finite-blocklength inequalities
\begin{equation}
 2k_1+k_2\le2n+2\log_2\frac1{1-2\epsilon_n},\qquad
 k_1+2k_2\le2n+2\log_2\frac1{1-2\epsilon_n}.
 \label{eq:poda-finite-open}
\end{equation}
If $\limsup\epsilon_n<1/2$, the additional terms are $O(1)$; division by $n$ gives the outer bound in \eqref{eq:poda-open-region}. The argument allows arbitrarily many predetermined switches. Its error threshold is $1/2$, and the asymptotic error must remain strictly below that threshold. The blockwise condition $\epsilon_n<1/2$ alone is insufficient: if $\epsilon_n\to1/2$, the additional terms in \eqref{eq:poda-finite-open} need not be $o(n)$.

\subsection{Three-Phase Open-Loop Coding}

Begin with the known pilot $U_1=U_2=1$. A receiver selecting port $s$ observes 1 if $H=s$ and 0 if $H=1-s$. An open-loop receiver uses this result only for decoding; its subsequent actions remain unchanged.

Let $a,b$ be user 1's two bits and $c,d$ user 2's. The three-slot code and its outputs are
\begin{center}
\begin{tabular}{lccc}
\toprule
Quantity & Slot 1 & Slot 2 & Slot 3\\
\midrule
$U_1$ & $a$ & $b$ & $b$\\
$U_2$ & $c$ & $c$ & $d$\\
$S_1$ & $1$ & $1$ & $0$\\
$S_2$ & $0$ & $1$ & $1$\\
\midrule
$Y_1$, $H=0$ & $a\oplus c$ & $b\oplus c$ & $b$\\
$Y_2$, $H=0$ & $c$ & $c\oplus b$ & $d\oplus b$\\
$Y_1$, $H=1$ & $a$ & $b$ & $b\oplus d$\\
$Y_2$, $H=1$ & $a\oplus c$ & $c$ & $d$\\
\bottomrule
\end{tabular}
\end{center}
For $H=0$, receiver 1 recovers $b,c,a$ in that order, while receiver 2 recovers $c,b,d$. For $H=1$, receiver 1 reads its bits from the first two outputs and receiver 2 from the last two. Each receiver thus recovers its own two bits with zero error from its local output alone.

Grouping the corresponding slots of $L$ such codes into three phases limits each receiver to one switch over the entire block. With $A,B,C,D\in\mathbb F_2^L$, transmit
\begin{align}
 U_1&=(A,B,B), &U_2&=(C,C,D),\\
 S_1&=(1^{2L},0^L), &S_2&=(0^L,1^{2L}).
 \label{eq:poda-one-switch}
\end{align}
During the pilot, receiver 1 selects port 1 and receiver 2 port 0, matching their first data actions. Each receiver switches only once. The block of length $3L+1$ carries $2L$ bits per user, approaching $(2/3,2/3)$.

The single-user points $(1,0)$ and $(0,1)$ are achieved by having the other transmitter send 0; the active user's output is then independent of the port and state. Append $u$ slots reserved for user 1 and $v$ slots reserved for user 2 after the three phases, keeping each receiver at its final port. The resulting block has length $3L+u+v+1$, carries $(2L+u,2L+v)$ bits, and requires no additional switch. Taking the closure gives the convex hull of
$(0,0),(1,0),(2/3,2/3),(0,1)$. Every code in this construction has zero error.

\subsection{Fixed-Port and Causal Capacity}

For any fixed ports $s_1,s_2$, there is a state in which at least one receiver observes $U_1^n\oplus U_2^n$ throughout the block. If the ports coincide, choose the state in which both receivers see this sum; if they differ, each state has one receiver that sees it.

In that state, the map from jointly correct message pairs to a suitable sum output is injective. If both receivers see the sum, their decoders determine both messages. If only receiver $i$ sees it, its correct decoder first determines $m_i$ and hence $U_i^n$. Subtraction gives $U_j^n$, from which the other, interference-free receiver's correct decoder determines $m_j$. The success probability in this state is therefore at most $2^n/(N_1N_2)$, while in the other state it is at most 1. Hence
\begin{equation}
 \epsilon_n\ge\frac12\left(1-\frac{2^n}{N_1N_2}\right),\qquad
 k_1+k_2\le n+\log_2\frac1{1-2\epsilon_n}.
 \label{eq:poda-finite-fixed}
\end{equation}
This proves $R_1+R_2\le1$. Time division, with the inactive user silent, achieves the entire triangle with zero error, without pilots or switching.

Under a causal policy, each receiver identifies $H$ independently from one known pilot and then selects $S_{i,t}=H$. During the data phase, each receives one interference-free bit of its own user per slot. A block of length $n$ carries $n-1$ bits per user with at most one switch.

Conversely, once a receiver's random seed is fixed, its actions are functions of previous outputs and do not increase the number of distinguishable output histories. A receiver has at most $2^n$ binary output histories, so its correct-decoding probability is at most $2^n/N_i$. Consequently,
\begin{equation}
 k_i\le n+\log_2\frac1{1-\epsilon_n},\qquad i=1,2.
 \label{eq:poda-active-cut}
\end{equation}
Thus $R_i\le1$. Together with the constructions above, this completes the proof of Theorem 1.

For $\epsilon\ge1/2$, fix both receivers at port 0 and send one pilot to identify the state. When $H=0$, each user then transmits one bit per data slot and both receivers decode directly. When $H=1$, the block may fail. The joint error is at most $1/2$, each user's rate approaches 1, and no switching is needed. For every $\epsilon<1$, \eqref{eq:poda-active-cut} bounds each user's rate by 1, so all three regions are the unit square. The restriction $\epsilon<1$ excludes the degenerate criterion that permits every transmission to fail.

\section{Capacity Bounds for the Erasure Interference Channel}

Now let the selected-port output be retained with probability $p$ and otherwise replaced by a recognizable erasure symbol. Erasures are independent of the inputs, state, actions, and past; $p=1$ recovers the unerased channel. The causal capacity region can still be determined exactly. For open-loop observation, we obtain inner and outer bounds.

\medskip\noindent
\textbf{Proposition 2 (Causal capacity with erasures).}
The causal capacity region of the erasure model is $[0,p]^2$. At most one switch per receiver per block remains sufficient. This holds for every fixed joint error threshold $\epsilon<1$.

\begin{proof}
For $0<p<1$, send $\ell_n\to\infty$ known pilots with $\ell_n=o(n)$, holding each receiver at a fixed port during training. Any unerased pilot identifies $H$; the probability of seeing none is $(1-p)^{\ell_n}$. After training, each receiver switches at most once to the estimated port. The two data links are then binary erasure channels. For binary random linear codes, the full-rank probability derived below implies achievability of every rate $R<p$: with probability tending to 1, the number of unerased rows exceeds the message dimension by a quantity linear in $n$, and the probability of rank deficiency tends to zero. The case $p=0$ gives the zero region, while $p=1$ requires one pilot.

For the converse, fix all random seeds and one receiver's erasure pattern. If $T_i$ positions are unerased, there are at most $2^{T_i}$ distinguishable output histories, even under causal actions. Fix also the other user's message and $H$. Distinct correctly decoded messages of user $i$ must then produce distinct histories. Hence
\[
 \Pr\{\widehat M_i=M_i\}
 \le\mathbb E\left[\min\left\{1,\frac{2^{T_i}}{N_i}\right\}\right],
 \qquad T_i\sim\operatorname{Bin}(n,p).
\]
If $R_i>p$, binomial concentration makes this upper bound tend to zero. Every fixed nontrivial error threshold therefore requires $R_i\le p$.
\end{proof}

For open-loop policies under vanishing joint error,
\begin{equation}
 p\mathcal C_{\rm ol}(1)\ \subseteq\ \mathcal C_{\rm ol}(p)
 \ \subseteq\
 \left\{\begin{array}{l}
 (R_1,R_2)\ge0,\quad R_i\le p,\\
 2R_1+pR_2\le2p,\quad pR_1+2R_2\le2p
 \end{array}\right\}.
 \label{eq:poda-erasure-bounds}
\end{equation}
For the inner bound, use the three-phase construction and encode $A,B,C,D$ with a common binary linear erasure code of rate slightly below $p$. Each required sum modulo two is also a codeword of this linear code. A receiver first recovers the phase codewords and then applies the three-slot algebra. Use $\ell_n\to\infty$ fixed-port pilots with $\ell_n=o(n)$ solely to identify $H$ at the final decoder. The three-phase actions are predetermined and do not depend on the pilot outcomes. The pilot ports and phase order can be chosen to require at most one switch per receiver.

For the outer bound, fix an open-loop sequence and couple the erasure patterns of a receiver in the two hypothetical states to a common Bernoulli-$p$ retention set $E$. This coupling preserves the marginal channel in each state. Let $A$ denote the interference-free positions in one state. If a uniform message is recovered from a record with error probability $e$, introducing the error indicator and separating correct and incorrect decoding gives
$H(M\mid\text{record})\le h_2(e)+e\log_2N$. Vanishing error in the equiprobable mixture implies vanishing error in each state. Since the output bound already bounds each user's rate, this conditional entropy is $o(n)$. Adding the information bounds for the two states yields
\begin{align*}
 2k_i\le 2pn
 &-H(U_{j,A\cap E}\mid E)
 -H(U_{j,A^{\rm c}\cap E}\mid E)+o(n)\\
 \le 2pn&-H(U_{j,E}\mid E)+o(n).
\end{align*}
Independent retention satisfies $H(U_{j,E}\mid E)\ge pH(U_j^n)$. To see this, expand by the chain rule: conditioning only on the retained past coordinates gives no smaller conditional entropy than conditioning on every past coordinate. Reliable decoding of the other user's message also gives
$H(U_j^n)\ge k_j-o(n)$, proving the two weighted outer bounds. For $p<1$, the inner and outer bounds generally differ; the exact open-loop and fixed-port regions remain undetermined.

\begin{figure}[htbp]
\centering\includegraphics[width=.70\textwidth]{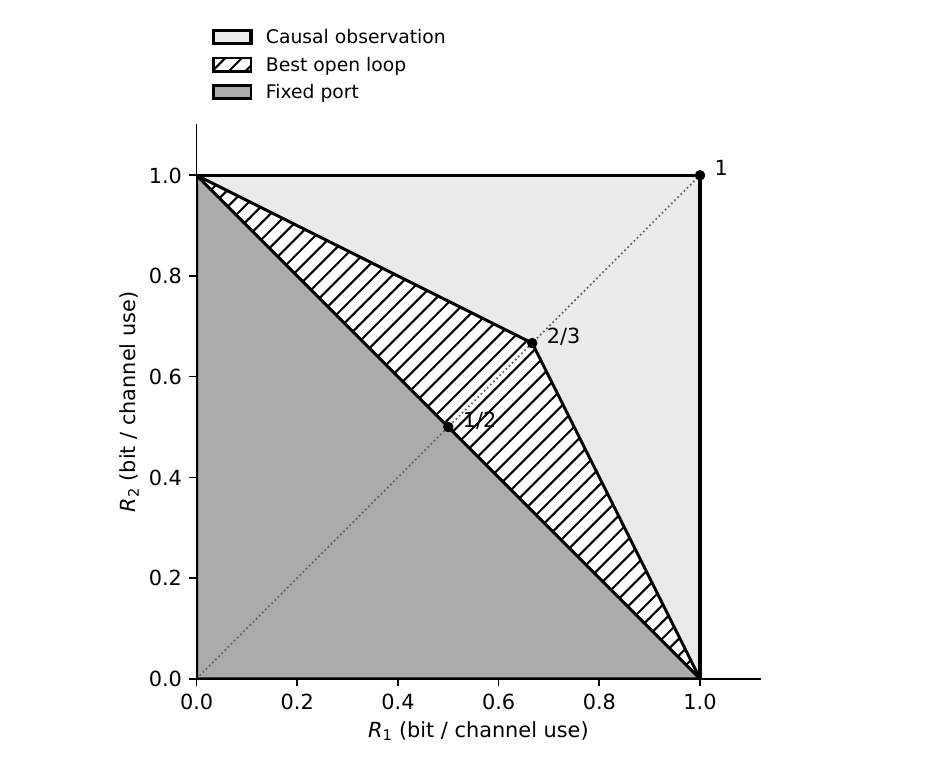}
\caption{The three exact capacity regions without erasures, for any fixed joint error threshold below $1/2$. Within the same port family, predetermined switching improves on fixed ports; causal observation further increases sum capacity from $4/3$ to 2.}
\label{fig:regions}
\end{figure}

\section{Port Identification and Coding Efficiency}
\subsection{Achievable Rates and Identification Overhead}

The open-loop construction combines port schedules with repeated symbols so that local records eliminate interference in either state. The causal construction first identifies the state and then selects the port that directly yields the desired symbol. The former exploits algebraic relations among records; the latter uses the history to choose subsequent channel sections.

The causal construction attains the output-trajectory counting bound. Three properties are used together: the port family contains an interference-free section, simultaneous known pilots identify its label, and the identification cost is negligible relative to blocklength. A single port switch suffices.

The zero-error constructions give, for a block of length $n$, $\lfloor n/2\rfloor$ bits per user under time division, $2\lfloor(n-1)/3\rfloor$ under the three-phase open-loop scheme, and $n-1$ under causal observation. At $n=192$, these payloads are 96, 126, and 191 bits. The symmetric zero-error open-loop outer bound requires $3k\le2n$, leaving a gap between the 126-bit construction and the finite-blocklength bound.


\section{Finite-Blocklength Pilot Allocation}

Although the pilot fraction may vanish in a capacity argument, pilot overhead directly limits the payload at finite blocklength. More pilots improve state identification but leave fewer slots for data. For pilot-assisted binary random linear codes, this tradeoff reduces to a finite integer optimization and yields an achievable guarantee for fixed codebooks.

\subsection{Pilot-Assisted Coding}

Retain the binary port $U_i\oplus(H\oplus S_i)U_j$ and the independent erasure model. In each slot, each receiver observes the exact selected bit with probability $p$ and an erasure otherwise. Erasures are mutually independent across receivers and time and are independent of the state, messages, codebooks, previous observations, and actions. Both ports have the same retention probability $p$. Switching in response to pilots therefore leaves the subsequent erasure law unchanged.

For blocklength $n$, choose in advance $m$ simultaneous pilot slots and $d=n-m$ data slots. During training, both transmitters send 1 and receiver $i$ holds a predetermined port $a_i$. Any unerased pilot identifies $H$, since the output is 1 exactly when $H=a_i$. If all pilots are erased, the receiver declares failure at the end of the block. Otherwise, it selects port $H$ at the start of the data phase and holds it thereafter. Each receiver thus switches at most once per block.

Each user carries $k$ bits $M_i\in\mathbb F_2^k$ and transmits $G_iM_i$ during the data phase, where $G_i$ is a $d\times k$ binary matrix. In the random coding ensemble, the matrices are independent, their entries are independent and uniform, and both matrices are independent of the state and erasures. Once selected, each matrix is known to the corresponding transmitter and receiver and is reused across blocks.

Let $E_i$ be the set of unerased data positions. After selecting the correct port $H$, the receiver observes coordinates of its own linear codeword. It decodes uniquely if $G_i[E_i,:]$ has full column rank and declares failure otherwise. The failure symbol lies outside the message set. Thus success occurs exactly when at least one pilot is received and the data equations have a unique solution; underdetermined systems are declared failures.

\subsection{A Random Linear Coding Bound}

\noindent\textbf{Lemma 2 (Full column rank over the binary field).}
An $r\times k$ binary matrix with independent uniform entries has full column rank with probability
\begin{equation}
 f(k,r)=
 \begin{cases}
  0,&r<k,\\[2pt]
  \displaystyle\prod_{j=0}^{k-1}(1-2^{j-r}),&r\ge k.
 \end{cases}
 \label{eq:poda-rank-probability}
\end{equation}
The empty product is 1, so $f(0,r)=1$.

\begin{proof}
Expose the columns successively. Each column is independent and uniform on $\mathbb F_2^r$. If the first $j$ columns are linearly independent, their span contains exactly $2^j$ vectors. The conditional probability that the next column lies outside this span is $1-2^j/2^r$. Multiplication over $j=0,\ldots,k-1$ proves the formula. If $r<k$, an $r$-dimensional space cannot contain $k$ linearly independent vectors, so the probability is zero.
\end{proof}

Define
\begin{align}
 q&=(1-p)^m,\label{eq:poda-pilot-erasure}\\
 a(d,k;p)&=\sum_{r=k}^{d}\binom dr p^r(1-p)^{d-r}f(k,r),
 \label{eq:poda-data-success}\\
 s(n,m,k;p)&=(1-q)a(n-m,k;p).
 \label{eq:poda-single-success}
\end{align}
Here $q$ is the probability that all pilots are erased at one receiver. Since the erasures and matrix are independent, conditioning on $r$ retained rows leaves their entries independent and uniform. Thus $a$ is the ensemble-average probability of unique data decoding, and $s$ is the ensemble-average success probability for one user.

\medskip\noindent
\textbf{Proposition 3 (Finite-blocklength pilot-assisted achievability).}
Under this protocol and its independence assumptions, the ensemble-average joint failure probability for $k$ bits per user is exactly
\begin{equation}
 B(n,m,k;p)=1-\bigl[s(n,m,k;p)\bigr]^2.
 \label{eq:poda-pilot-bound}
\end{equation}
There consequently exists a fixed pair of binary matrices $(G_1,G_2)$ whose joint failure probability is at most $B(n,m,k;p)$, with the same blocklength, pilot count, and one-switch constraint. The guarantee holds for both average and maximal message error.

\begin{proof}
Conditional on any state $H=h$, success at one receiver is equivalent to receiving at least one local pilot and obtaining a retained local matrix of full column rank. Pilot and data erasures are independent, so the probability is $(1-q)a=s$. Neither this event's characterization nor its probability depends on $h$.

The receivers' matrices and erasures are independent. Conditional on $H=h$, their success events are therefore independent and have joint probability $s^2$. Averaging over the common state leaves $s^2$ unchanged. This product form relies on conditional independence given $H$ and on success probabilities that do not depend on $H$. It does not apply to correlated erasures or to retention laws that vary with the state or history.

The joint failure probability averaged over the finite set of matrix pairs is $B$, so some pair has failure probability at most $B$. With the matrices fixed, failure depends only on the erasure patterns and the ranks of the retained submatrices, not on the messages. Every message pair therefore has the same error probability, and the average and maximal error guarantees coincide.
\end{proof}

\subsection{Error Allocation and Payload Optimization}

Write $a=a(n-m,k;p)$. Separating state-identification failure from data-decoding failure in \eqref{eq:poda-pilot-bound} gives
\begin{equation}
 B=(2q-q^2)+(1-q)^2(1-a^2).
 \label{eq:poda-error-budget}
\end{equation}
The first term is the probability that at least one receiver loses every pilot. The second is the probability that both receivers identify the state but unique data decoding fails. Increasing $m$ reduces $q$ and shortens the data phase $d=n-m$; their combined effect determines the optimum.

Since $a\le1$, the condition $B\le\epsilon$ requires
\begin{equation}
 q\le1-\sqrt{1-\epsilon}.
\end{equation}
For $0<p<1$ and $0<\epsilon<1$, the pilot length must therefore satisfy
\begin{equation}
 m\ge
 \left\lceil
 \frac{\ln(1-\sqrt{1-\epsilon})}{\ln(1-p)}
 \right\rceil.
 \label{eq:poda-minimum-pilots}
\end{equation}
Equation~\eqref{eq:poda-minimum-pilots} is a necessary lower bound on the pilot count. Payload feasibility still requires the full constraint, including data-decoding failure.

For given $n,p,\epsilon$, define the symmetric message size guaranteed by this construction as
\begin{equation}
 \begin{split}
 K_{\rm pilot\text{-}lin}(n,p,\epsilon)=\max\bigl(&\{0\}\ \cup\\
 &\{k\in\mathbb Z_{\ge1}:\exists m\in\{1,\ldots,n-k\},\;
 B(n,m,k;p)\le\epsilon\}\bigr).
 \end{split}
 \label{eq:poda-pilot-optimization}
\end{equation}
Choose a maximizing $m^*$. Proposition 3 guarantees a fixed codebook pair supporting
$K_{\rm pilot\text{-}lin}/n$ bits per slot per user with joint failure probability at most $\epsilon$. The denominator includes the pilots. If the feasible set contains no positive integer, the payload is zero; no pilots are then needed and the message error is zero.

The optimization is solved by searching the finite set of pilot counts. For fixed $m$, the functions $f(k,r)$, $a$, and $s$ are nonincreasing in $k$, so $B$ is nondecreasing. A binary search therefore finds the largest feasible payload for each $m$. This optimum concerns the achievable bound of the specified ensemble; a particular fixed codebook may have a smaller error probability. For example, when $p=1$, rank-deficient matrices give the random ensemble a positive average failure probability, whereas choosing a full-rank matrix permits zero-error communication.

For unequal payloads, the same argument gives the fixed-codebook achievability bound
$1-s(n,m,k_1;p)s(n,m,k_2;p)$. Equation~\eqref{eq:poda-pilot-optimization} specializes to the symmetric design $k_1=k_2$.

\subsection{Finite-Blocklength Results}

Set $n=192$, $p=0.9$, and $\epsilon=0.01$. Exhaustive evaluation of all 18,336 feasible positive integer pairs $(m,k)$ in \eqref{eq:poda-pilot-optimization} gives
\begin{equation}
 m^*=3,\qquad K_{\rm pilot\text{-}lin}=156,
 \qquad B(192,3,156;0.9)=0.00931617.
\end{equation}
Hence a fixed codebook pair exists with 156 bits per user, nominal sum rate 1.625 bits per slot, and joint failure probability below 1\%. For this ensemble-average bound, three pilots uniquely attain the maximum payload.

\begin{figure}[htbp]
\centering\includegraphics[width=.94\textwidth]{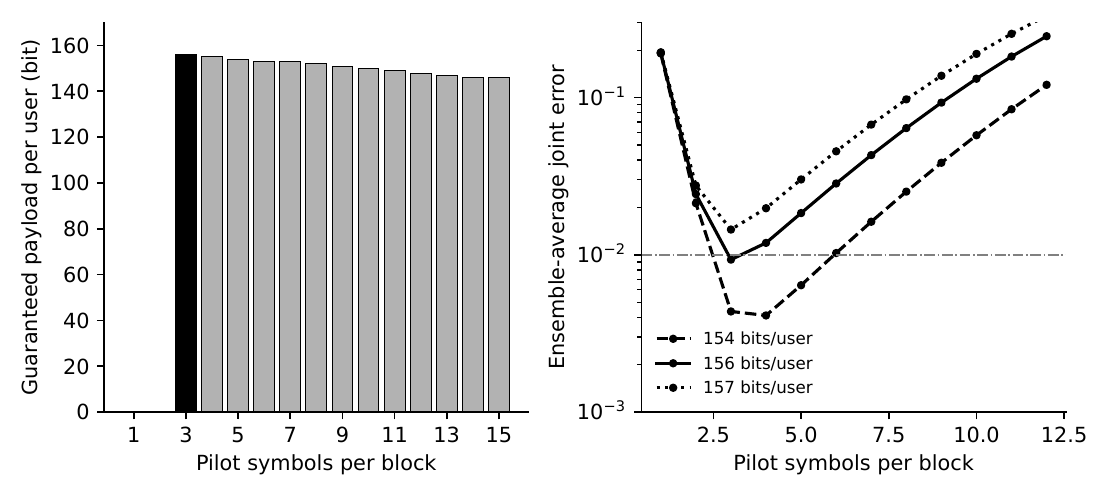}
\caption{Allocation between state identification and data redundancy. Left: largest payload satisfying the exact achievable bound for each pilot count. One and two pilots admit no positive payload at the target error and are plotted as zero. Right: ensemble-average failure probability at fixed payloads; the horizontal line marks 1\%.}
\label{fig:pilot-opt}
\end{figure}

With fewer than three pilots, state-identification failure alone exceeds the target: two pilots give a joint identification failure probability of 1.99\%. Beyond three pilots, the loss of data redundancy becomes dominant. Four pilots give ensemble-average failure probability 0.01190838 at a payload of 156 bits, reducing the largest payload admitted by the bound to 155 bits. With three pilots and 157 bits, the failure probability rises to 0.01447296. Figure~\ref{fig:pilot-opt} shows the tradeoff.

The causal code matrices fixed in advance for the experiment below were also evaluated at 156 bits per user over 5,000 independent blocks. There were 52 joint failures, giving a point estimate of 1.04\% and a single-configuration one-sided 95\% upper confidence bound of 1.309\%. Because this upper bound exceeds the 1\% threshold, the main experiment uses the 154-bit payload that passed simultaneous confidence validation. These finite-sample observations are consistent with the existence guarantee at 156 bits and do not establish that the tested matrices' true failure probability exceeds 1\%.

\section{Coding Performance on the Binary Erasure Model}
\subsection{Recoverability of the Desired Message}

The users independently transmit uniform $k$-bit messages. Their code matrices are generated and published in advance and remain fixed throughout the experiment. From its erasure indicators and identified state, receiver $i$ collects binary equations
\begin{equation}
 y_i=A_i m_i+B_i m_j,\qquad j\ne i.
 \label{eq:linear-record}
\end{equation}
Unique recovery of $m_i$ requires precisely
\begin{equation}
 d_i=\rank[A_i\ B_i]-\rank B_i=k.
 \label{eq:desired-rank}
\end{equation}
To see this, quotient the observation space by the column space of $B_i$. The image of $A_i$ in this quotient has dimension $d_i$; the map from the desired message to that image is injective if and only if $d_i=k$. Full column rank of the joint matrix is a stronger condition than~\eqref{eq:desired-rank}.

For independent uniform messages, the unresolved number of desired-message degrees of freedom is $k-d_i$ for each realization of the state, actions, and erasure pattern. Averaging gives
\begin{equation}
 H(M_i\mid Y_i^n,S_i^n,H,E_i^n,V_i)=k-\E[d_i].
 \label{eq:rank-entropy}
\end{equation}
In this experiment, actions depend only on known pilots and erasure indicators and are independent of the data messages. For data-dependent actions, the posterior also includes constraints imposed by action selection and must be recomputed.

When all pilots are erased, the receiver cannot identify $H$ and declares a local failure rather than guessing the state. Offline calculations may still use the true $H$ on these blocks, but~\eqref{eq:rank-entropy} then measures entropy with the state supplied as additional information. The separately reported undelivered payload, $kP_{e,i}$, is a different quantity.

The implementation uses binary Gaussian elimination, eliminating interference columns first and then back-substituting on the desired-message columns. Each trial generates messages, transmits codewords, forms the right-hand-side observations, and compares the decoded messages with those sent. Elimination and back-substitution are performed in every recovery; the rank condition determines uniqueness. The main experiment contains 354240 blocks and verifies 539949 successful local message recoveries, with no undetected errors observed. A decoder that declares failure does not guess a message, so failed payloads are distinguished from bit errors in delivered payloads.

\subsection{Coding Schemes and Statistical Methods}

The main experiment uses $n=192$ and $p=0.9$. Schemes requiring state identification use three simultaneous known pilots, leaving 189 data slots. Six schemes are compared:
\begin{enumerate}
\item \emph{Time division.} Each user occupies 96 slots while the other transmits zero. Each uses a fixed random linear code, with no probing or switching.
\item \emph{Fixed-port coding.} Both receivers always use port 0. Pilots are used only for state identification at decoding; data use independent dense binary codes.
\item \emph{Predetermined scanning.} Ports 0 and 1 each occupy approximately half the data phase, according to a fixed schedule. Dense binary codes are used.
\item \emph{Structured open-loop coding.} The three-phase arrangement in~\eqref{eq:poda-one-switch} maps each user's message linearly into two length-63 vectors and repeats the required phase. The pilot port equals the first data-phase port, giving one switch over the block.
\item \emph{Causal observation.} After the three pilots, each receiver identifies $H$ locally and keeps $S_i=H$ during data transmission, with at most one switch.
\item \emph{Known-state reference.} Each receiver knows $H$ before transmission, and all 192 slots carry data. This ideal reference has zero identification cost and a different initial information set from the first five schemes.
\end{enumerate}

The first five schemes have the same transmission budget, observation budget, and port family. Time division needs no pilots and is therefore not charged an artificial identification cost. Fixed-port, scanning, and causal schemes use their own fixed code matrices; random seeds are not selected using the simulation outcomes. Structured and dense code families differ in coding design rather than in unaccounted transmission resources.

Screening and validation samples are independent. A coarse payload grid uses 320 blocks per point, followed by an integer grid with 640 blocks. Three candidate payloads per scheme are then frozen and each tested on a further 5000 blocks. One-sided exact binomial upper bounds have simultaneous coverage at least 95\% across all 18 candidates. For $f$ observed joint failures, the upper bound $u$ solves
\begin{equation}
 \sum_{j=0}^{f}\binom{5000}{j}u^j(1-u)^{5000-j}
 =\frac{0.05}{18}.
 \label{eq:confidence}
\end{equation}
The table selects the largest candidate with $u\le0.01$. This is the largest validated payload among the specified matrices and candidates.

Let $P_{e,i}$ denote the block failure probability of user $i$, and let $P_e$ be the joint failure probability. Define the nominal sum rate, goodput, and joint-delivery goodput by
\begin{equation}
 R_\Sigma=\frac{2k}{n},\qquad
 G_\Sigma=\frac{k}{n}(2-P_{e,1}-P_{e,2}),\qquad
 G_{\rm joint}=\frac{2k}{n}(1-P_e).
\end{equation}
The first counts the payload assigned to failed blocks as well; the second counts each successful user's payload; the third counts a block only when both users succeed. Percentage comparisons below primarily concern $G_\Sigma$.

\begin{table}[H]
\centering\small\setlength{\tabcolsep}{3.5pt}
\caption{Payload and goodput at a fixed blocklength. The confidence bounds cover all 18 prespecified configurations simultaneously.}
\label{tab:main}
\begin{tabular}{lrrrrr}
\toprule
Policy & \shortstack{Payload\\per user} & Sum rate & Goodput & \shortstack{Joint\\error} & \shortstack{95\%\\bound}\\
 & bit & bit/use & bit/use & \% & \%\\
\midrule
Time division & 74 & 0.7708 & 0.7692 & 0.42 & 0.744\\
Fixed port & 77 & 0.8021 & 0.7998 & 0.56 & 0.922\\
Open-loop scan & 77 & 0.8021 & 0.7998 & 0.56 & 0.922\\
Structured open loop & 93 & 0.9688 & 0.9660 & 0.56 & 0.922\\
Causal observation & 154 & 1.6042 & 1.5997 & 0.56 & 0.922\\
Known state & 158 & 1.6458 & 1.6411 & 0.58 & 0.947\\
\bottomrule
\end{tabular}
\end{table}
\begin{table}[H]
\centering\small\setlength{\tabcolsep}{3.5pt}
\caption{Resource use and decoding statistics at the selected payloads, averaged over the two receivers. The first decodable slot is averaged over successful blocks only.}
\label{tab:resources}
\begin{tabular}{lrrrrr}
\toprule
Policy & Pilots & Switches & First decode & Row XORs & bit/read\\
\midrule
Time division & 0 & 0.000 & 131.9 & 1770 & 0.3846\\
Fixed port & 3 & 0.000 & 133.6 & 6053 & 0.3999\\
Open-loop scan & 3 & 1.000 & 134.5 & 3783 & 0.3999\\
Structured open loop & 3 & 1.000 & 157.8 & 4133 & 0.4830\\
Causal observation & 3 & 0.506 & 175.9 & 7053 & 0.7998\\
Known state & 0 & 0.000 & 177.3 & 7293 & 0.8205\\
\bottomrule
\end{tabular}
\end{table}
\begin{table}[H]
\centering\small\setlength{\tabcolsep}{3.5pt}
\caption{Independent comparison at a common payload of 80 bits per user, with 5000 blocks per policy. Goodput is in bit/use. Residual entropy is conditioned additionally on the true state.}
\label{tab:common}
\begin{tabular}{lrrrr}
\toprule
Policy & \shortstack{Joint error\\(\%)} & Goodput & \shortstack{Residual entropy\\(bit)} & Row XORs\\
\midrule
Time division & 11.90 & 0.7823 & 0.0790 & 1791\\
Fixed port & 2.78 & 0.8214 & 0.0186 & 6119\\
Open-loop scan & 4.14 & 0.8160 & 0.0233 & 3702\\
Structured open loop & 0.26 & 0.8323 & 0.0000 & 3996\\
Causal observation & 0.26 & 0.8323 & 0.0000 & 5155\\
Known state & 0.00 & 0.8333 & 0.0000 & 5268\\
\bottomrule
\end{tabular}
\end{table}

\subsection{Goodput, Reliability, and Computational Cost}

In Table~\ref{tab:main}, causal observation supports 154 bits per user and structured open-loop coding supports 93. Their goodputs are 1.5997 and 0.9660 bits/slot, respectively, an increase of 65.6\%. In this configuration, causal observation improves goodput by 100\% over fixed-port coding and predetermined scanning, and by about 108.0\% over time division. The 65.6\% figure compares the tested code families at the stated erasure probability and blocklength; the 50\% sum-capacity increase without erasures follows separately from the asymptotic coding theorem.

Causal goodput is about 2.52\% below the known-state reference, reflecting both pilot occupancy and pilot loss. The probability that all three pilots are erased at a receiver is $10^{-3}$, and the probability that this occurs at at least one receiver is $1-(1-10^{-3})^2=0.001999$. This consumes part of the 1\% joint error budget, requiring a joint allocation of pilot length and data redundancy.

Table~\ref{tab:resources} shows that causal observation uses the same number of observations and about 0.506 switches on average, compared with one for predetermined scanning. At the selected payload, each receiver performs about 7053 row-XOR operations, versus 4133 for the structured open-loop scheme, largely because it carries and recovers more information. These counts include elimination only, excluding encoding, final back-substitution, and precomputation. They are zero when all pilots are lost and decoding stops. All schemes have a scheduled block delay of 192 slots. The first decodable time, reported only on successful blocks, is an additional statistic rather than a delay including retransmissions.

Each block uses 384 receiver observations, so information delivered per observation is $G_\Sigma/2$. Charging only for binary transmission slots gives the same value per unit of total transmission cost. These quantify symbol and observation efficiency; energy efficiency requires a physical energy model. The fairness index of the two users' mean goodputs,
$\mathcal J=(g_1+g_2)^2/[2(g_1^2+g_2^2)]$, is close to one, consistent with the symmetric traffic and channel setting.

\begin{figure}[htbp]
\centering\includegraphics[width=.94\textwidth]{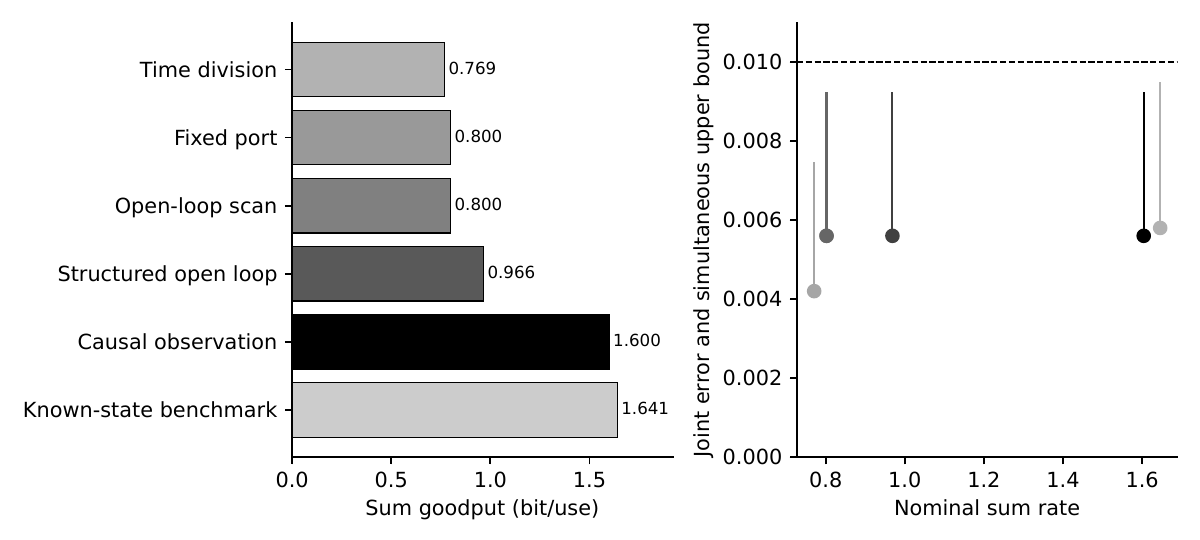}
\caption{Validated goodput and joint failure probability. The upper ends of the vertical lines in the right panel are one-sided 95\% bounds with simultaneous coverage across 18 candidates; the dashed line marks the 1\% target. The known-state scheme is an ideal reference.}
\end{figure}

\subsection{Equal-Payload Performance and Erasure Sensitivity}

In addition to the largest candidate payload under the reliability constraint, Table~\ref{tab:common} compares failure rates and computational cost at 80 bits per user. Each configuration uses an independent set of 5000 blocks. Equal payloads separate the benefit of observation design from the additional decoding work associated with carrying more information. Zero observed failures are interpreted through the corresponding confidence upper bound.

\begin{figure}[htbp]
\centering\includegraphics[width=.89\textwidth]{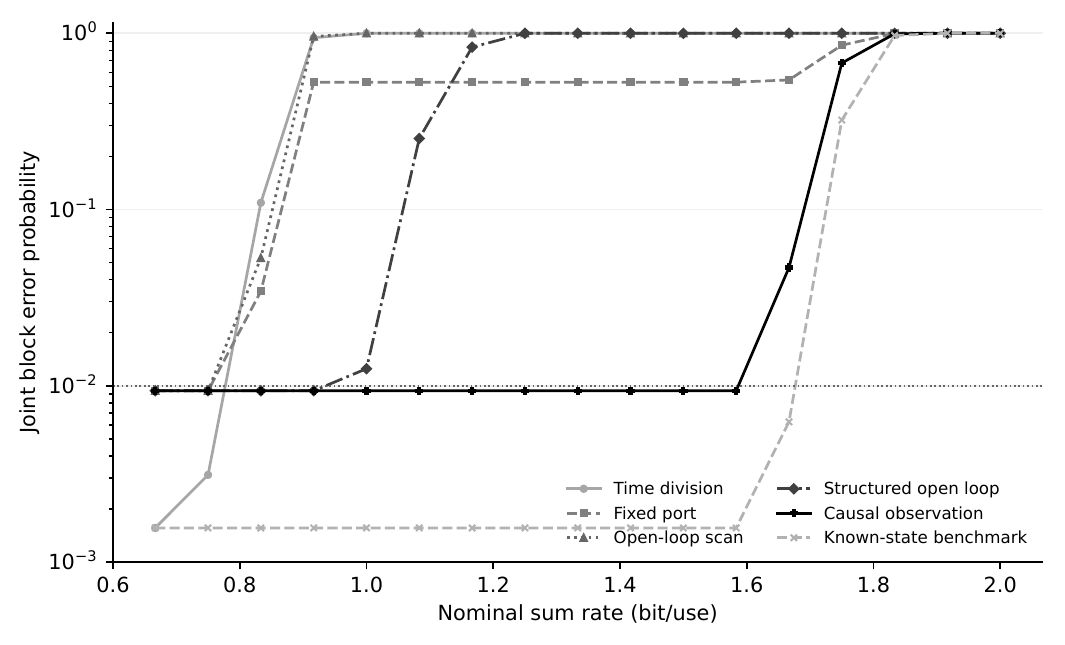}
\caption{Payload sweeps for fixed code matrices, with 320 blocks per point. Zero-failure points are plotted at $0.5/320$. The final comparison uses the independent validation in Table~\ref{tab:main}.}
\label{fig:curves}
\end{figure}

\begin{figure}[htbp]
\centering\includegraphics[width=\textwidth]{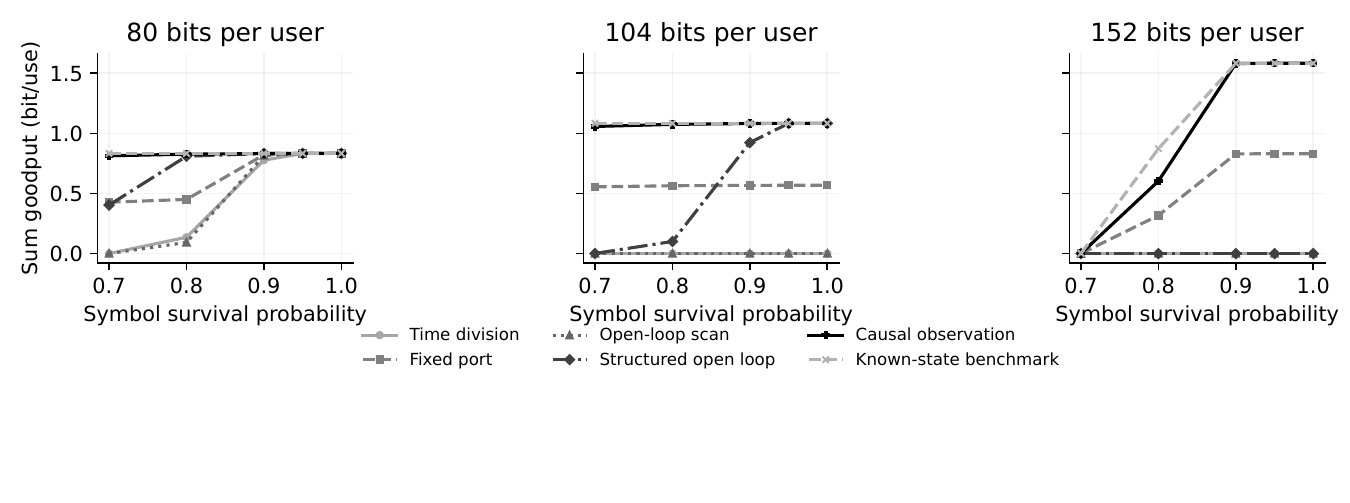}
\caption{Erasure sensitivity at three fixed payloads, with 1600 blocks per point. The same fixed code matrices are used throughout.}
\label{fig:sweep}
\end{figure}

Figure~\ref{fig:curves} shows coarse-grid failure curves for the different code families; these are screening results rather than the final validation data. Figure~\ref{fig:sweep} holds the payload fixed and varies the symbol retention probability. At higher payloads, successful port identification cannot compensate for too few retained symbols. Selecting the matched port removes the interference in this model, while coding redundancy is still needed for independent erasures.


\FloatBarrier
\section{Conclusion}

Causal observation achieves the interference-free capacity region of the binary two-port interference channel studied here. For every fixed joint error threshold below $1/2$, the exact fixed-port, open-loop, and causal sum capacities are $1$, $4/3$, and $2$ bits per channel use. The 100\% gain over optimal fixed-port operation and the 50\% gain over optimal open-loop operation follow from matching upper bounds and achievable constructions. They require neither transmitter feedback nor receiver cooperation, and one port switch per receiver is sufficient. Independent erasures preserve the conclusion in the form $\mathcal C_{\rm act}=[0,p]^2$.

The physical--observation dual-axis description makes the source of this gain explicit. A known pilot reveals which formation kernel separates the desired symbol from the interfering symbol. The causal policy then uses that kernel for the remaining transmission. Since identifying a block-constant state costs a vanishing fraction of a long block, both users recover their full interference-free rates. In this model, the observation policy is therefore part of the capacity-determining communication protocol.

The finite-horizon results give a corresponding design principle for fixed codes. The joint posterior and remaining resource budget determine which observation should be acquired next. Exact recursion in the finite model and second-order approximation bounds for continuous directions make the optimization quantitative. The array and coding experiments then account for estimation error, pilot occupancy, switching, and decoding cost. Their gains depend on the task: a policy with higher mutual information need not minimize message error, and higher whole-frame goodput need not imply a lower bit error rate.

The longstanding interference-capacity question motivates this work; the theorem established here concerns a channel with receiver-selectable complementary ports and an identifiable block state. General prescribed interference channels, continuous models without a clean feasible direction, and the exact open-loop erasure region require further results. The present contribution is an exact capacity separation under explicit observation resources, together with methods for optimizing the observations themselves. It provides a concrete observer-mediated coding theorem in which causal control of the observation axis removes the full interference penalty.

\appendix
\section{Numerical Methods and Reproduction}

The source archive contains the manuscript, vector figures, raw results, fixed code matrices, the C++ decoding kernel for the binary model, and Python implementations of observation design and coding experiments. The principal commands are
\begingroup\small
\begin{verbatim}
python3 simulation/continuous_observation_solver.py --summarize
python3 simulation/verify_continuous_tree_enumeration.py
python3 simulation/verify_continuous_kernel.py
python3 simulation/verify_continuous_interpolation.py
python3 make_continuous_figures.py
python3 simulation/finite_observation_optimum.py
python3 simulation/robust_global_optimum.py
python3 simulation/dynamic_observation.py
python3 make_dynamic_figures.py
python3 simulation/observation_design.py
python3 simulation/verify_observation_design.py
python3 make_observation_figures.py
python3 simulation/run.py
python3 simulation/pilot_tradeoff.py
python3 make_figures.py
bash build.sh
\end{verbatim}
\endgroup
The continuous-policy summary command uses saved results; the README gives the commands for a fresh direction search. Compile the manuscript twice with XeLaTeX. Code matrices and stage-specific random seeds are fixed. Screening, validation, equal-payload comparisons, and parameter sweeps are stored separately so that the data used to select a configuration are not reused to validate its performance.

Reproduction uses the stated probability laws and published code matrices, with no external data dependency.

The three-slot erasure-free construction is checked exhaustively over all binary messages and both states. An independent Python equation checker reconstructs the binary experiment slot by slot, verifies decoded messages, and checks the one-switch constraint. Separate checks cover closed-form combining weights, attainment of the worst-case response errors, orthogonal training, and agreement between the likelihood implementation and direct enumeration. The robust-weight program checks every stationary-point candidate. Finite observation policies use integer probability recursions and are also compared with exhaustive trees at short horizons. Dynamic array experiments form the scalar observations at every slot and retain the pilot schedule and independent screening and validation data. These checks concern the finite implementations. Confidence intervals quantify sampling uncertainty within the stated model and code; they do not include unmodeled jumps, synchronization errors, or hardware distortion.
\begingroup\small\setstretch{1.02}

\endgroup
\end{document}